\documentclass[10pt,twocolumn]{article}

\usepackage[utf8]{inputenc}
\usepackage[T1]{fontenc}
\usepackage[english]{babel}
\usepackage{amsmath,amssymb,amsfonts}
\usepackage{graphicx}
\usepackage{physics}
\usepackage{hyperref}
\usepackage{cite}
\usepackage{authblk}
\usepackage{geometry}
\usepackage{xcolor}
\usepackage{caption}
\usepackage{subcaption}
\usepackage{appendix}
\usepackage{float}
\usepackage{dblfloatfix}

\usepackage{comment}
\usepackage{tabularx}
\usepackage[font=small,labelfont=bf]{caption}
\usepackage{comment}
\usepackage{xcolor} 
\usepackage{balance}
\usepackage{etoolbox}
\renewcommand{\refname}{}
\makeatletter
\patchcmd{\thebibliography}{\section*{\refname}}{}{}{}
\patchcmd{\thebibliography}{\section*{\bibname}}{}{}{}
\makeatother

\newcommand{\mg}[1]{\textcolor{blue}{\textbf{[MG:} #1\textbf{]}}}

\newcommand{\myparagraph}[1]{%
  \vspace{0pt}\noindent\textit{#1.}\hspace{0pt}%
}

\title{\textbf{Maximum Independent Set via Probabilistic and \\ Quantum Cellular Automata}}

\author[1]{Federico Dell'Anna}
\author[1,2]{Matteo Grotti}
\author[1]{Vito Giardinelli}
\affil[1]{\textit{Dipartimento di Fisica e Astronomia dell’Università di Bologna, I-40127 Bologna, Italy}}
\affil[2]{\textit{INFN, Sezione di Bologna, I-40127 Bologna, Italy}}

\begin{document}
\newtheorem{conj}{Conjecture}
\twocolumn[
\begin{@twocolumnfalse}
\date{}
\maketitle
\begin{abstract}
We study probabilistic cellular automata (PCA) and quantum cellular automata (QCA) as frameworks for solving the Maximum Independent Set (MIS) problem. We first introduce a synchronous PCA whose dynamics drives the system toward the manifold of maximal independent sets. Numerical evidence shows that the MIS convergence probability increases significantly as the activation probability $p \to 1$, and we characterize how the steps required to reach the absorbing state scale with system size and graph connectivity.
Motivated by this behavior, we construct a QCA combining a pure dissipative phase with a constraint-preserving unitary evolution that redistributes probability within this manifold. Tensor Network simulations reveal that repeated dissipative--unitary cycles concentrate population on MIS configurations. We also provide an empirical estimate of how the convergence time scales with graph size, suggesting that QCA dynamics can provide an efficient alternative to adiabatic and variational quantum optimization methods based exclusively on local and translationally invariant rules.

\end{abstract}
\vspace{0.5cm}
\end{@twocolumnfalse}
]

\section{Introduction}

Cellular automata (CA) provide a versatile framework for modelling complex dynamical systems through parallel local update rules acting on a lattice \cite{Bhattacharjee2016,Chopard1998,BilottaPantano2010}. Quantum Cellular Automata (QCA) extend cellular automata to the quantum domain, implementing completely positive trace-preserving maps acting on local subsystems and providing an intrinsically distributed architecture for computation and simulation, characterised by strict locality, parallelism, and translational invariance \cite{Farrelly2020_QCA_review,ArrighiNesmeWerner2007_1DQCA}. Both unitary and non-unitary versions of QCA have been theoretically formulated and experimentally realised, for example using Rydberg-atom platforms \cite{Unitary_Wintermantel}. These systems exhibit rich dynamics, ranging from coherent evolution to dissipative many-body processes \cite{BonebergCarolloLesanovsky2023_DissipativeQCA,wagner2024density} and non-unitary information flow \cite{WagnerNigmatullinGilchristBrennen2022_InfoFlowNonUnitaryQCA}. Such flexibility has led to increasing interest in embedding computation within QCA dynamics, enabling both reversible logic and open-system protocols within physically grounded architectures \cite{GillmanCarolloLesanovsky2023_QCA_QNN}.

In this work, we explore this potential in the context of a paradigmatic NP-hard optimisation problem: the Maximum Independent Set (MIS) \cite{robson1986algorithms,das2012heuristics,gupta2021simple}. Given a graph $G$, the MIS consists of the largest subset of vertices with no pair connected by an edge. MIS is central to combinatorial optimisation and graph theory \cite{GareyJohnson1979,NemhauserWolsey1988}, with applications spanning network design, resource allocation, scheduling, and constraint satisfaction \cite{schuetz2022combinatorial}. Classical approaches include exact exponential-time algorithms \cite{tarjan1977finding} as well as approximation and parameterised schemes for restricted graph classes such as unit-disk graphs \cite{NiebergHassinVanLeeuwen2004,DasFonsecaJallu2016,Andrist2023_HardnessMISUDG}. 
In quantum computation, MIS is typically encoded in the ground state of an Ising-type Hamiltonian \cite{Choi2010_MIS_Hamiltonian,Lucas2014_IsingNPproblems}, whose minimisation can be addressed by adiabatic schemes or variational algorithms \cite{Choi2020_MIS_HamiltonianGap}. Recent works have exploited this formulation to implement MIS computation using Rydberg atom arrays, showing both theoretical prospects for quantum advantage and experimental feasibility on near-term devices \cite{Ebadi2022_MIS_Rydberg,Serret2020_RydbergAnalogQA,Andrist2023_HardnessMISUDG}.
These Hamiltonian encodings have therefore become standard benchmarks for exploring potential quantum advantage in combinatorial optimisation.

Here, we pursue a complementary route leveraging the intrinsic structure of cellular automata. We first introduce a classical \emph{probabilistic cellular automaton} (PCA) designed to generate maximal independent sets. The proposed PCA belongs to the class of stochastic local-update rules: its dynamics enforce the independence constraint by resolving local conflicts, while sites with no occupied neighbours are promoted to the occupied state with probability $p$.
As a consequence, the PCA induces a Markov chain on the configuration space whose stationary states coincide exactly with maximal independent sets. This rule is closely related to classical stochastic optimisation methods, including randomized greedy algorithms \cite{luby,Greedy_MIS} and Glauber dynamics for the hard-core lattice gas model \cite{Glauber,glauber_mc}, even though they do not provide clear estimates concerning the probability to detect a MIS of the graph. 

Guided by this stochastic formulation, we extend the construction to a quantum cellular automaton that combines dissipative and unitary evolutions. In contrast with global Hamiltonian optimisation, our QCA approach exploits locality and parallelism at every step: dissipative maps enforce the independence constraint, while coherent unitary updates generate superpositions that facilitate exploration of multiple candidate solutions.

The paper is organised as follows. Section~\ref{sec:MIS} introduces the MIS problem through both its discrete and Hamiltonian formulations. Section~\ref{sec:pca} defines probabilistic cellular automata and presents our PCA-based MIS solver. Section~\ref{sec:QCA} extends the method to quantum cellular automata. Section~\ref{sec:classical_simulation} analyses classical simulation results, while Section~\ref{sec:quantum_simulation} presents the quantum simulations. Finally, Section~\ref{sec:conclusion} provides an outlook and concludes the work.

\section{The problem}
\label{sec:MIS}

Let \(G=(V,E)\) be a finite undirected graph with vertex set \(V\) and edge set \(E\).
An \emph{independent set} \(S\subseteq V\) is a subset of vertices in which no two elements are adjacent in \(G\). The \emph{Maximum Independent Set (MIS)} problem consists in finding an independent set of largest possible size, whose cardinality is denoted by \(|\mathrm{MIS}(G)|\) \cite{tarjan1977finding}.

More formally: 
\[
S(G) = \left\{ \mathbf{s} \in \{0,1\}^{|V|}
\;\middle|
s_i + s_j \le 1 \;\forall (i,j) \in E
\right\},
\]
and therefore
\[
\mathrm{MIS}(G) = \arg\max_{\mathbf{s} \in S(G)} \sum_{i \in V} s_i .
\]
where each decision variable \(s_i\in\{0,1\}\) encodes whether the vertex \(i\) is included (\(s_i=1\)) in the selected independent set \(S\) \cite{NemhauserWolsey1988,GareyJohnson1979}.

In general, an independent set of $G$ is said to be \emph{maximal} when no additional node can enter the independent set without violating the no-adjacency condition. Graphs may have exponentially many maximal independent sets, as shown in \cite{Nielsen2002_NumberMIS,PalmerPatkos2022_SurveyMIS}, whereas recent bounds indicate that the number of maximum independent sets is much more limited \cite{LevitItskovich2025_MaximumISBounds}. This supports the general observation that most graphs have only few maximum independent sets compared to their maximal independent sets.

Equivalently, another formulation of the problem maps the integer variables \(s_i\in\{0,1\}\) to a cost function whose ground state encodes the MIS. The usual Hamiltonian formulation is:
\begin{equation}\label{eq:H-MIS}
H \;=\; -\sum_{i\in V} s_i \;+\; U \sum_{(i,j)\in E} s_i s_j,
\end{equation}
with penalty strength \(U>0\) \cite{Choi2010_MIS_Hamiltonian
,Lucas2014_IsingNPproblems
,Choi2020_MIS_HamiltonianGap}. The first term rewards inclusion of vertices; the second penalizes occupied adjacent pairs. For \(U>1\) the ground state of
\(\;H\;\) corresponds to a maximum independent set (up to the
unimportant constant energy offset) because any edge with both endpoints occupied is energetically disfavored relative to occupying a single endpoint. This formulation is often used when mapping the MIS to quantum Hamiltonians. \\

\section{Probabilistic Cellular Automata}
\label{sec:pca}
Probabilistic cellular automata (PCA) are spatially extended, discrete-time stochastic dynamical systems defined on a lattice $\Lambda$~\cite{AgapieAndreicaGiuclea2014_PCA,BusicMairesseMarcovic2010_PCAPerfectSampling,CasseMarcovic2020_PCAmemoryTwo,BayraktarLuMaggioniWuYang2024_PCAlocalTransition}. 
Each lattice site (or cell) $i\in\Lambda$ carries a discrete state $s_i(n)\in\mathcal{S}$ (typically $\mathcal{S}=\{0,1\}$), and the configuration at discrete time $n$ is denoted by $\mathbf{s}(n) = (s_i(n))_{i\in\Lambda}$. 
The system evolves according to a synchronous local randomized update rule: at each time step, the new state of site $i$ is drawn from a probability distribution that depends only on the states in a finite neighbourhood $\mathcal{N}_i \subset \Lambda$.

More precisely, let $\Theta\big(\xi \,\big|\, s_{\mathcal{N}_i}\big)$ be the probability that the new state of site $i$ becomes $\xi \in \mathcal{S}$ given the neighbourhood configuration $s_{\mathcal{N}_i} \in \mathcal{S}^{\mathcal{N}_i}$. 
Under synchronous updates, the global transition rule is the product of the local ones, so that the PCA induces a Markov chain on the configuration space $\mathcal{S}^{\Lambda}$ with one-step transition kernel
\begin{equation}
\mathcal{P}(\mathbf{s}' \mid \mathbf{s})
=
\prod_{i \in \Lambda}
\Theta\big(s_i' \mid s_{\mathcal{N}_i}\big),
\end{equation}
where $s_{\mathcal{N}_i}$ denotes the restriction of $\mathbf{s}$ to the neighbourhood of site $i$. \\

\begin{figure}
    \centering
\includegraphics[width=0.99\linewidth]{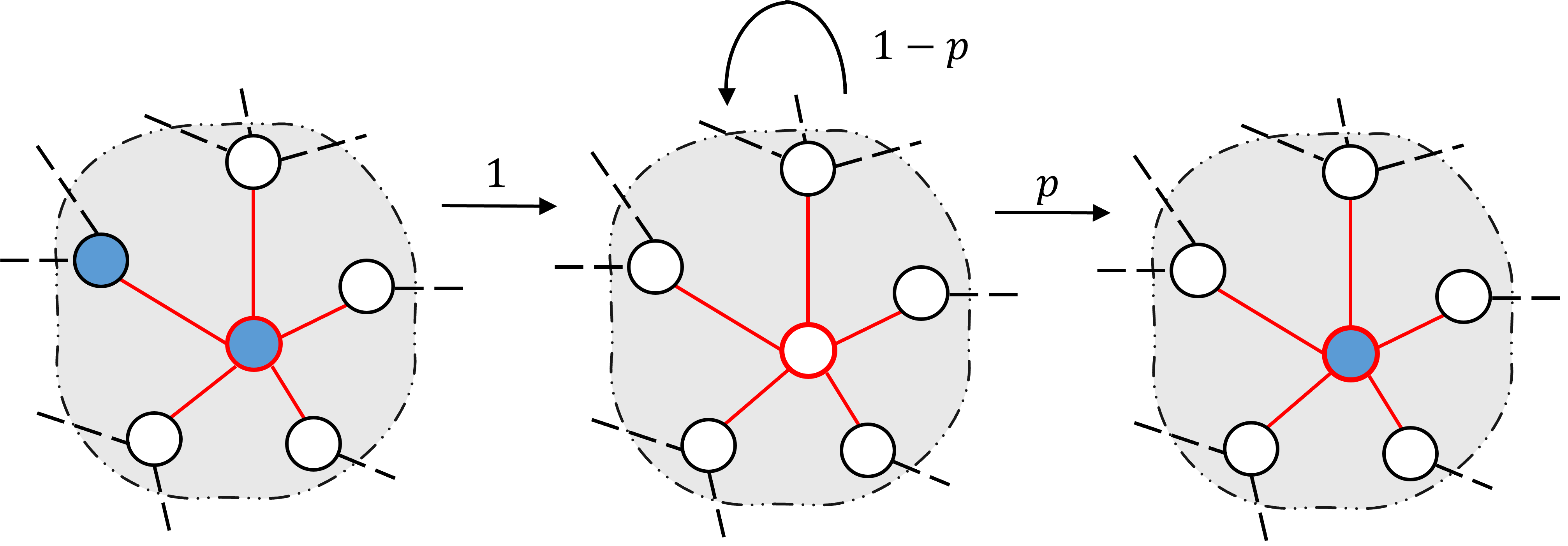}
\caption{Illustration of the probabilistic local update rule: if at least one neighbour is active, the cell becomes inactive with probability $1$ (left). If all neighbours are inactive and the cell is inactive, it may remains inactive (center) or it activates (right) with probability $1-p$ and $p$ respectively.}
    \label{fig:transition_prob}
\end{figure}

\myparagraph{\textbf{PCA for MIS problem}} 
We introduce a PCA designed to have the maximal independent sets of a graph $G$ as the absorbing states. 
Each vertex of $G$ is associated with a cell of the automaton, and neighbour relations in the graph define the interaction neighbourhoods of each cell. 
Cell states take values in $\mathcal{S}=\{0,1\}$, where state $1$ represents an \emph{active} node. 
The dynamics start from the uniform configuration $s_i(0)=0$ for all $i \in \Lambda$.
The local update rule at discrete time $n$ is defined as follows (see Fig.~\ref{fig:transition_prob}):
\begin{itemize}
\item[\(\triangleright\)] if at least one neighbour of cell $i$ is active, i.e., $\exists\, j\in\mathcal{N}_i : s_j(n)=1$, then $i$ becomes inactive at the next step regardless of its current state:
$s_i(n+1) = 0$;
\item[\(\triangleright\)] if all neighbours of cell $i$ are inactive, i.e., $s_j(n)=0\ \forall j\in\mathcal{N}_i$, then: \\
- if $s_i(n)=1$, the state is preserved: $s_i(n+1)=1$. \\
- if $s_i(n)=0$, the cell activates with probability $p$:
\begin{equation}
s_i(n+1) =
\begin{cases}
1 & \text{with probability } p,\\
0 & \text{with probability } 1-p.
\end{cases}
\end{equation}
\end{itemize}
By construction, the dynamics eventually converge to a configuration that is guaranteed to be a maximal independent set. 
The independence constraint is automatically enforced because any two adjacent active cells are both deactivated at the next iteration. 
Furthermore, maximality follows from the fact that any inactive cell surrounded by inactive neighbours will eventually become active with non-zero probability.
Hence, the set of \emph{stationary states}, also called \emph{absorbing states}, of the Markov chain generated by this PCA coincides exactly with the set of maximal independent sets of $G$. The algorithm we propose shares conceptual similarities to stochastic methods based on Glauber dynamics \cite{Glauber}: both frameworks employ probabilistic rules to construct maximal independent sets. However, a fundamental difference separates the two approaches: the standard Glauber dynamics updates one vertex at a time (asynchronously) and therefore never introduces conflicts between adjacent vertices. This behaviour typically results in faster convergence to stationarity compared to our synchronous update rule, which may temporarily activate adjacent vertices and subsequently resolve the conflicts. Nevertheless, the synchronous update mechanism enables us to provide an estimate of the probability that our algorithm converges to a maximum independent set (MIS) across different graph sizes and connectivity regimes. In particular, such probability increases monotonically with the parameter $p$ (see Fig.~\ref{fig:tp_classical}). By contrast, Glauber-based methods offer evidence of fast convergence times, estimated to be $O(N \log (N k))$ \cite{Fast_glauber}, where $N$ denotes the number of nodes and $k$ the maximum degree. However, if a high probability of convergence to a MIS is required, these guarantees hold only within a limited range of connectivity.

From our simulations aimed at assessing the performance of this PCA (see Sec. \ref{sec:classical_simulation}), we consistently observed the following behavior, which we therefore put forward as a conjecture: \emph{in the limit $p \to 1$, the probability of converging to any MIS configuration is greater than to the probability of converging to any mIS configuration.}

In other words, we conjecture that, in the regime of large $p$, the algorithm tends to favor convergence to maximal independent sets of higher cardinality. This effect can be understood by noting that a node can be activated only when \emph{all} its nearest neighbors are inactive. For a node with many neighbors, the probability that all of them simultaneously sit in state $0$ decreases exponentially with its degree, meaning that highly connected nodes typically activate much later than low-degree nodes. As a result, the dynamics preferentially activates low-degree nodes first; once activated, these nodes exclude only a small portion of the graph from further activation. This mechanism allows a larger number of nodes to become active overall, thus biasing the PCA dynamics toward maximal independent sets with larger cardinality.

The MIS convergence probability can be computed once the graph topology is known, as illustrated in Appendix~\ref{app:calcolo_pmis} for a minimal example. Although in most simulated cases the MIS convergence probability approaches 1 as $p \to 1$, we also provide a counterexample showing that this behaviour does not hold in general.


\section{Quantum dynamics}
\label{sec:QCA}
We consider a quantum protocol combining dissipative and unitary dynamics to explore the space of independent-set configurations on a generic graph $G = (V,E)$ of $N = |V|$ sites. 
The evolution is described in terms of two Liouvillian generators, $\mathcal{L}_1$ and $\mathcal{L}_2$, corresponding respectively to a purely dissipative and unitary process. 
Starting from the vacuum configuration $|0\rangle^{\otimes N}$, the overall dynamics consists of an initial dissipative relaxation followed by the repeated alternation of these two processes. \\

The dynamics of an open quantum system is governed by the Lindblad master equation
\begin{equation}
\frac{d\hat{\rho}(t)}{dt} 
= 
\mathcal{L}(\hat{\rho})
= 
-i[\hat{H},\hat{\rho}]
+ \sum_l 
\left(
\hat{L}_l \hat{\rho} \hat{L}_l^\dagger 
- \tfrac{1}{2}\{\hat{L}_l^\dagger \hat{L}_l, \hat{\rho}\}
\right),
\label{eq:lindblad}
\end{equation}
where $\{\cdot,\cdot\}$ denotes the anticommutator and $\hat{H}$ is the system Hamiltonian.

To study the steady-state convergence of the Liouvillian, it is convenient to employ the vectorization procedure, which maps density matrices to state vectors $\hat{\rho} \to |\rho\rangle\rangle$ and superoperators to ordinary operators $\mathcal{L} \to \hat{\mathcal{L}}$. 
Following the standard construction \cite{Gilchrist2009_VectorizationQuantumOperations},
\begin{align}
\hat{\mathcal{L}} &=
-i\,\hat{H} \otimes \hat{\mathbb{I}} 
+ i\,\hat{\mathbb{I}} \otimes \hat{H}^T
\nonumber\\
&\quad
+ \sum_l 
\hat{L}_l \otimes \hat{L}_l^{\dagger\,T}
-\frac{1}{2}\hat{L}_l^\dagger \hat{L}_l \otimes \hat{\mathbb{I}}
-\frac{1}{2}\hat{\mathbb{I}} \otimes (\hat{L}_l^\dagger \hat{L}_l)^T.
\label{eq:vectorized}
\end{align}
For two density matrices $\hat{\rho}$ and $\hat{\mu}$ written in vectorized form, 
$|\rho\rangle\rangle$ and $|\mu\rangle\rangle$, 
it is possible to define the generalized scalar product
\begin{equation}
    \mathrm{Tr}[\hat{\rho}^\dagger \hat{\mu}]
= \langle\!\langle \rho | \mu \rangle\!\rangle.
\label{scalar_product}
\end{equation}
\myparagraph{\textbf{Dissipative dynamics}} The first, purely dissipative, evolution is generated by a Liouvillian $\hat{\mathcal{L}}_1$ with $\hat{H}=0$, where the jump operators $\hat{L}_l$ implement stochastic activation and deactivation processes on the graph.
For each node $i$ with neighborhood $\mathcal{N}_i$, we define local operators:
\begin{equation}
\hat{\sigma}^- = |0\rangle\langle 1|,\;
\hat{\sigma}^+ = |1\rangle\langle 0|,\;
\hat{P}^0 = |0\rangle\langle 0|,\;
\hat{P}^1 = |1\rangle\langle 1|.
\end{equation}

The activation process ($0 \to 1$) occurs only if all neighboring nodes are in the state $|0\rangle$, and is represented by
\begin{equation}
\hat{L}^{(i)}_{\text{on}}
= 
\hat{\sigma}^+_i 
\prod_{j \in \mathcal{N}_i} \hat{P}^0_{j}.
\end{equation}

Conversely, the deactivation process ($1 \to 0$) is triggered whenever at least one neighboring node is excited.  
For each node $i$, we define a family of jump operators corresponding to all possible local configurations of its neighborhood:
\begin{equation}
\hat{L}^{(i,\mathbf{c})}_{\text{off}}
=
\hat{\sigma}^-_i 
\prod_{j \in \mathcal{N}_i} 
\big[
  (1 - c_j)\,\hat{P}^0_{j} + c_j\,\hat{P}^1_{j}
\big],
\end{equation}

Here, the binary vector $\mathbf{c}=(c_1,c_2, ..., c_j, ...)$ encodes one of the $2^{|\mathcal{N}_i|}-1$ allowed neighborhood configurations, namely, all possible combinations of ground and excited neighbors, except the fully unexcited one.  
The complete dissipative generator can then be compactly written as
\begin{equation}
\big\{
  \hat{L}^{(i)}_{\text{on}},
  \hat{L}^{(i,\mathbf{c})}_{\text{off}}
  \,\big|\,
  i \in V,\;
  \mathbf{c} \in \{0,1\}^{|\mathcal{N}_i|}\setminus\{(0,\ldots,0)\}
\big\}.
\end{equation}

This dissipative map fills the graph with excitations wherever the constraints allow, driving the system into a statistical mixture of all maximal independent sets.

In fact, starting from the vacuum configuration $|00\ldots0\rangle$, the Lindbladian dynamics progressively excites individual sites according to the local activation rules. 
At each infinitesimal time step, all possible jump processes, corresponding to the allowed local transitions, act incoherently and in parallel, effectively exploring the space of admissible configurations. 
The dynamics thus fills the graph with excitations in the all possible ways until no further activation or deactivation process can occur. 
As a result, the steady states of this dissipative evolution correspond precisely to the \emph{maximal independent sets} of the underlying graph, 
i.e., configurations that are locally stable under all the defined jump operators. 
Among these, configurations with the highest number of excitations, the \emph{maximum independent sets}, naturally acquire the largest steady-state populations if the initial state is the vacuum configuration $|00\ldots0\rangle$. 
This occurs because the dissipative dynamics tends to populate configurations with higher occupation numbers, as these can be reached through a greater number of allowed excitation pathways. 
Consequently, the stationary probabilities of the maximal configurations decrease monotonically with their cardinality.

\myparagraph{\textbf{Unitary evolution}}
Another evolution used in the protocol is a coherent, constraint-preserving one governed by the Hamiltonian
\begin{equation}
\hat{H}_{\mathrm{PXP}} 
= 
\sum_{i \in V}
\bigg(
  \prod_{j \in \mathcal{N}_i} \hat{P}^0_{j}
\bigg)
\hat{\sigma}^x_{i},
\end{equation}
where $\hat{\sigma}^x_{i}$ is the Pauli-$x$ operator acting on site $i$. 
This operator acts locally only when all neighbors of $i$ are in the ground state, ensuring that the exclusion rule is preserved at all times. 
The name ``PXP'' is chosen by analogy with the well-known one-dimensional PXP Hamiltonian, which is recovered exactly when the underlying graph $G$ is a simple open chain.
The PXP Hamiltonian thus defines a coherent dynamics restricted to the independent-set subspace, allowing reversible transitions between locally valid configurations.

\myparagraph{\textbf{Full QCA protocol}} The complete quantum cellular automaton protocol can be expressed as the sequential application of the dissipative and unitary evolutions. 
In the vectorized picture, starting from the vacuum configuration 
$|0\rangle\rangle^{\otimes N}$, 
the total evolution after $r$ cycles reads
\begin{equation}
| \psi(t) \rangle\!\rangle =
\underbrace{
e^{\hat{\mathcal{L}}_1 t}
e^{\hat{\mathcal{L}}_2 \theta}
\cdots
e^{\hat{\mathcal{L}}_1 t}
e^{\hat{\mathcal{L}}_2 \theta}
}_{\text{$r$-times}}
\;
e^{\hat{\mathcal{L}}_1 t}
|0\rangle\rangle^{\otimes N}.
\label{eq:full_evolution}
\end{equation}
where $\hat{\mathcal{L}}_2(\hat{\rho})  =
-i\,\hat{H}_{PXP} \otimes \hat{\mathbb{I}} 
+ i\,\hat{\mathbb{I}} \otimes \hat{H}^T_{PXP}$ 
denotes the unitary evolution associated with the PXP Hamiltonian. 
A first dissipative stage $e^{\hat{\mathcal{L}}_1 t}
|0\rangle\rangle^{\otimes N}$ projects the system onto the manifold of maximal independent sets, while the subsequent alternation of coherent and dissipative steps iteratively redistributes probability within this manifold, gradually favoring configurations of larger cardinality.
After each coherent step, the state can be written as:
\begin{equation}
|\psi\rangle
=
\sum_{M} c_M |M\rangle
+
\sum_{m} c_m |m\rangle
+
\sum_{i} c_i |i\rangle
+
c_0 |0\ldots 0\rangle,
\end{equation}
where $|M\rangle$ denotes maximum independent sets, $|m\rangle$ maximal but non-maximum independent sets, $|i\rangle$ independent but non-maximal configurations, and $|0\ldots0\rangle$ the vacuum state.
The subsequent dissipative evolution acts as a projection onto the manifold of maximal independent sets: it leaves $|M\rangle$ and $|m\rangle$ invariant (they are stationary states of all jump operators), while it drives any  independent configuration $|i\rangle$ towards some maximal configuration, through successive allowed activation processes. 

In particular, as argued above, when the dynamics starts from the state $\ket{00\ldots 0}$, the dissipative process tends to favor the MIS configurations. However, there is no theoretical reason to expect the same behavior for a generic state $\ket{i}$, since the probability of converging to an MIS depends on the specific activation paths accessible from such configuration. Nevertheless, across all graph instances considered in this work, we consistently observe that, for every value of $r$, the maximum independent sets acquire higher steady-state populations, leading us to conclude that these states remain favored even when the dissipative process acts on generic $\ket{i}$.

This protocol can be realistically implemented on multifrequency addressing Rybderg atoms. 
In Appendix~\ref{app:rydberg}, we show how this can be achieved by extending the QCA protocol in~\cite{Unitary_Wintermantel} to two-dimensional systems.

\section{Classical simulation}
\label{sec:classical_simulation}
To evaluate the performance of the proposed algorithm, we employ the following key metrics:
\begin{itemize}
\item \emph{MIS convergence probability} $P_{\text{MIS}}$, defined as the probability that the PCA state converges to a Maximum Independent Set (MIS). Consequently, the probability that the system converges to a maximal independent set (mIS) of smaller cardinality is $P_{\text{mIS}} = 1 - P_{\text{MIS}}$.
\item \emph{Convergence number of steps} $n$, defined as the number of steps required for the algorithm to reach a stationary configuration, i.e., a mIS of the graph.
\end{itemize}

To characterize the graph instances, we consider two complexity metrics: the \emph{number of nodes} $N$ and the \emph{average node degree} $k$, defined as the average degree over all nodes of the graph.

\paragraph{$\mathbf{P_{\text{MIS}}}$ \textit{heatmap}.}
For the classical simulations, we generated 10 random graph instances for each $(N, k)$ pair considered:
{\small
\begin{align*}
N=10 &, \quad k=2.0 \\
N=14 &, \quad k \in \{2.0, 2.5\} \\
N=18 &, \quad k \in \{2.0, 2.5, 3.0\} \\
N=22 &, \quad k \in \{2.0, 2.5, 3.0, 3.5\} \\
N=26 &, \quad k \in \{2.0, 2.5, 3.0, 3.5, 4.0\}
\end{align*}
}
The increasing range of $k$ with $N$ reflects the fact that, for larger graphs, higher levels of overall connectivity are possible, allowing for the generation of more diverse and structurally complex instances.

\begin{figure}[] 
     \centering
     \begin{subfigure}[b]{0.24\textwidth}
         \centering
         \includegraphics[width=\textwidth]{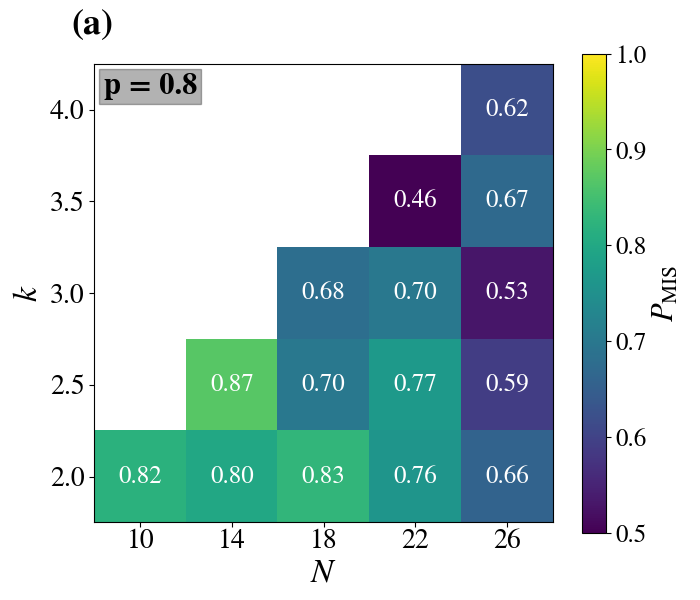}
         \label{}
     \end{subfigure}
     \begin{subfigure}[b]{0.24\textwidth}
         \centering
         \includegraphics[width=\textwidth]{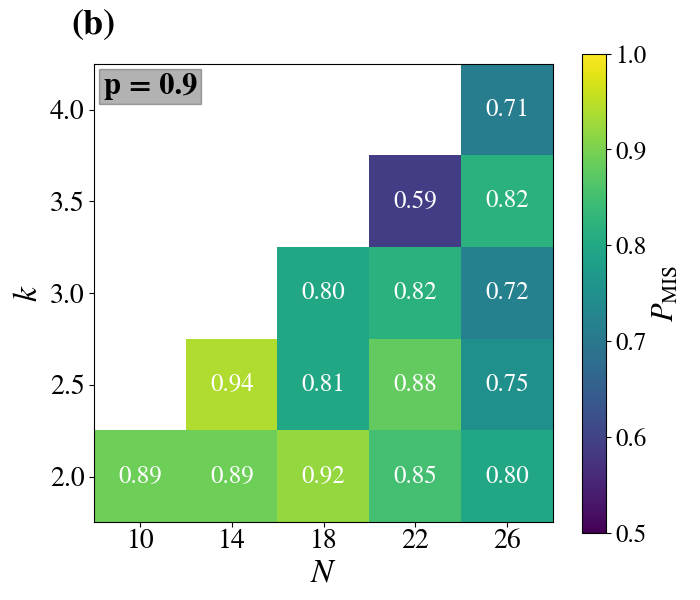}
         \label{}
     \end{subfigure}
     \hfill
     \begin{subfigure}[b]{0.24\textwidth}
         \centering
         \includegraphics[width=\textwidth]{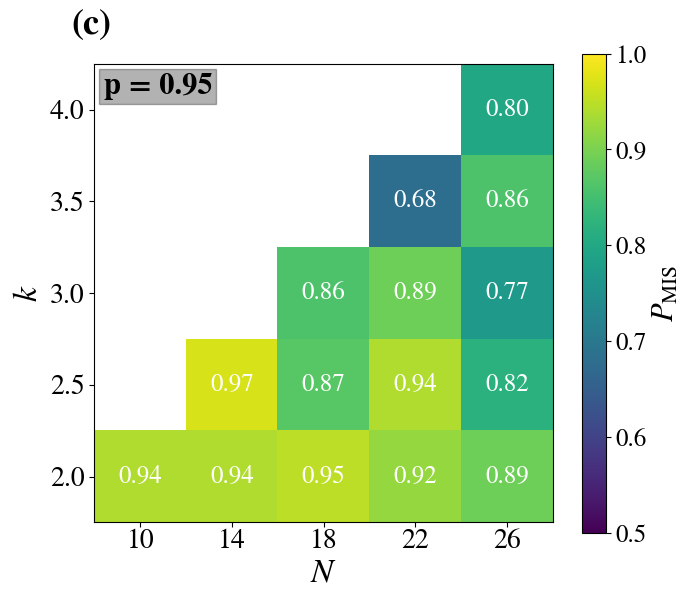}
         \label{}
     \end{subfigure}
     \begin{subfigure}[b]{0.24\textwidth}
         \centering
         \includegraphics[width=\textwidth]{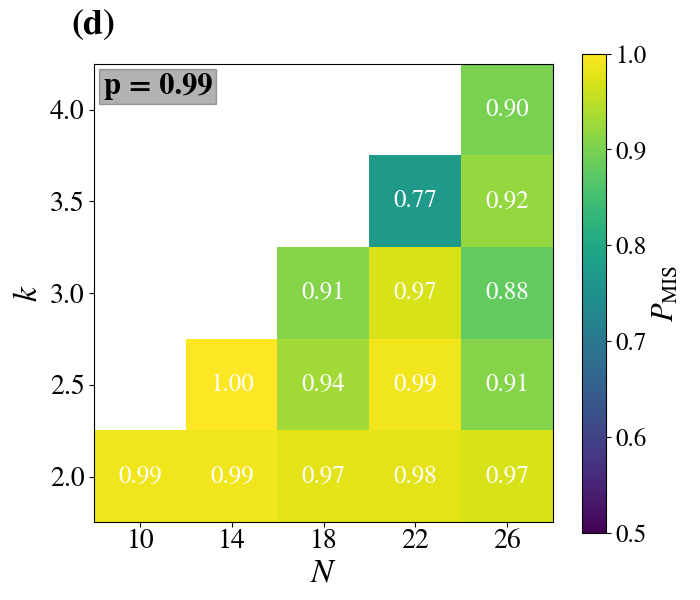}
         \label{}
     \end{subfigure}
     \caption{MIS convergence probability for different values of ($N$,$k$) and for the following values of $p$: 0.8 (a), 0.9 (b), 0.95 (c) and 0.99 (d). Each value is computed as the average on a set of 10 graph instances with same number of nodes $N$ and average degree $k$}
     \label{fig:heatmap_pmis_classical}
\end{figure}

For each graph instance, we performed 1000 independent runs of the algorithm up to convergence, using four different values of the transition probability $p~\in~\{0.8, 0.9, 0.95, 0.99\}$. 
For each value of $p$, we measured the corresponding $P_{\text{MIS}}$. We therefore computed, for each tuple ($N$,$k$,$p$), the average of $P_{\text{MIS}}$ over the set of 10 graph instances. The results are shown in Fig. \ref{fig:heatmap_pmis_classical}.

The comparison across panels shows that higher values of $p$ systematically increase $P_{\mathrm{MIS}}$, as the update rule more aggressively activates nodes and more efficiently suppresses incompatible pairs. 
However, this enhancement becomes progressively less effective as $N$ and $k$ grow: even for $p = 0.99$, the heatmap exhibits a clear fading of the MIS convergence probability in the upper-right region, indicating that the intrinsic combinatorial structure of the problem, rather than the local update strength, dominates in the large-size/high-connectivity regime.

\begin{figure}[h] 
     \centering
     \begin{subfigure}[b]{0.24\textwidth}
         \centering
         \includegraphics[width=\textwidth]{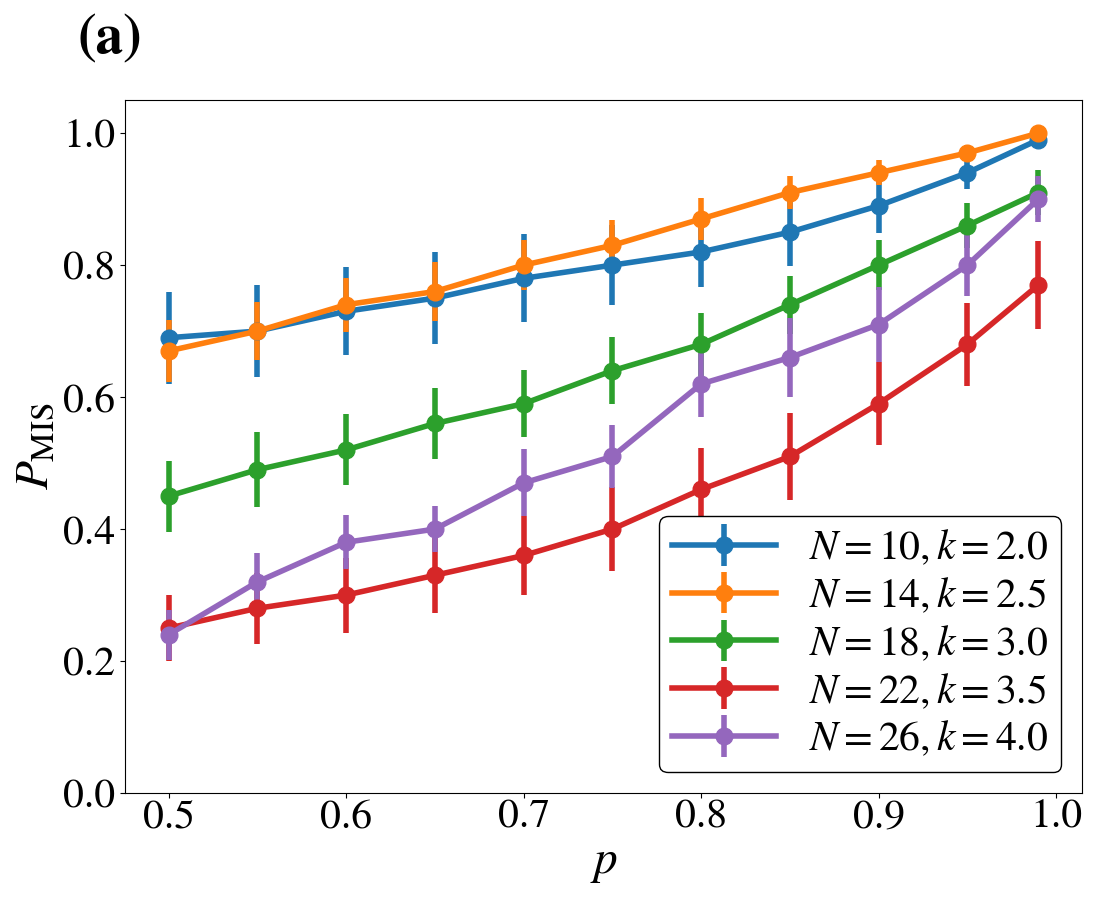}
         \label{}
     \end{subfigure}
     \begin{subfigure}[b]{0.24\textwidth}
         \centering
         \includegraphics[width=\textwidth]{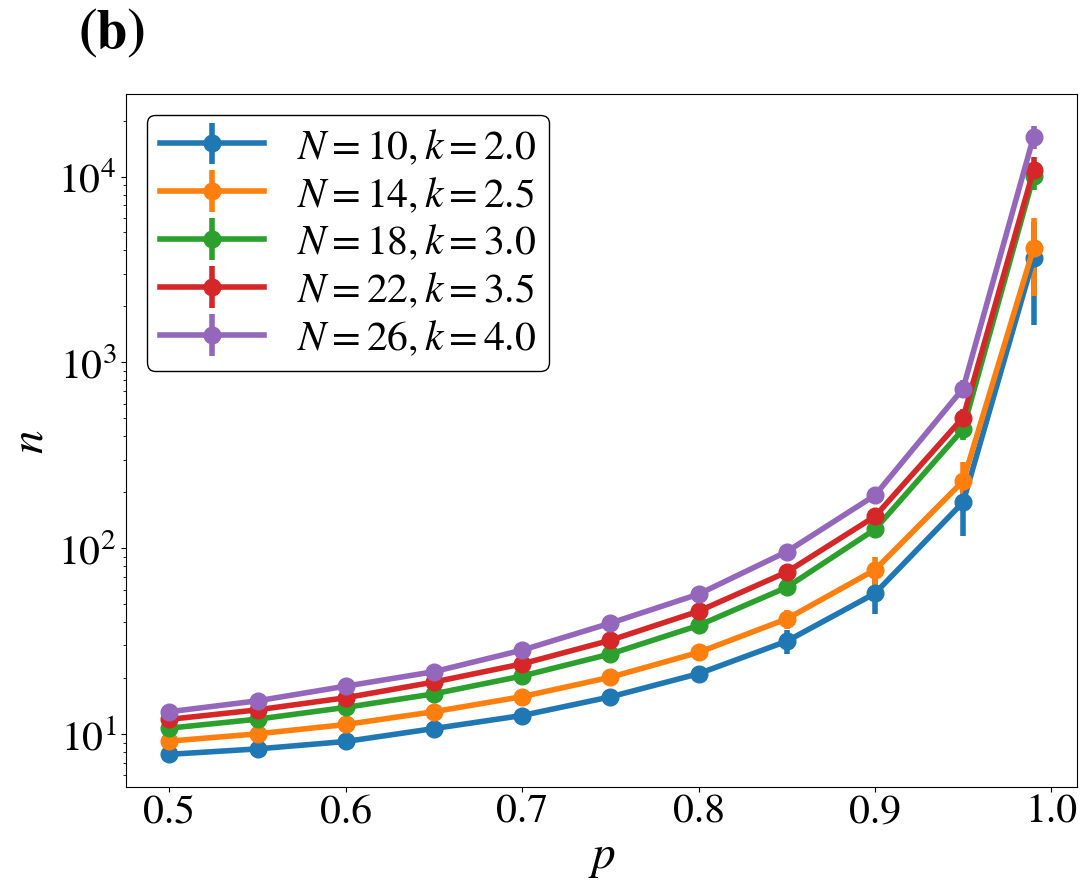}
         \label{}
     \end{subfigure}
     \caption{(a) MIS convergence probability $P_{\text{MIS}}$ (a) and convergence time (b) as a function of $p$ for different values of ($N$,$k$). For each ($N$,$k$) pair, we averaged $P_{\text{MIS}}$ and $n$ from a set of 10 graph instances, computed for different values of $p$}
     \label{fig:tp_classical}
\end{figure}

\paragraph{$\mathbf{P_{\text{MIS}}}$ \textit{scaling.}}
We next analyzed how the convergence probability $P_{\text{MIS}}$ and the convergence steps $n$ scale with the parameter $p$. To this end, we ran 1000 cycles of the algorithm for the same 10 graph instances used in the previous analysis, varying $p$ in the interval $[0.50, 0.99]$ with a step of $0.05$. 
Figure~\ref{fig:tp_classical} summarizes the scaling behavior of $P_{\text{MIS}}$ and $n$ as functions of the transition probability $p$ (panels (a) and (b), respectively).

These results bring into focus the fact that increasing $p$ systematically enhances the probability of convergence, but it also tends to increase the time needed to achieve a stationary solution. Therefore, by appropriately tuning the activation probability $p$, one can control the likelihood of reaching the optimal MIS configuration, balancing convergence speed and success probability.
\paragraph{\textit{Steps scaling}.}
Finally, we derived approximate scaling laws for the convergence steps as functions of $N$ and $k$, considered separately. Specifically, we performed 100 independent runs of the algorithm up to convergence, each on a randomly generated graph instance to ensure statistical independence, with varying numbers of nodes and connectivities.

\begin{figure}[h] 
     \centering
     \begin{subfigure}[b]{0.24\textwidth}
         \centering
         \includegraphics[width=\textwidth]{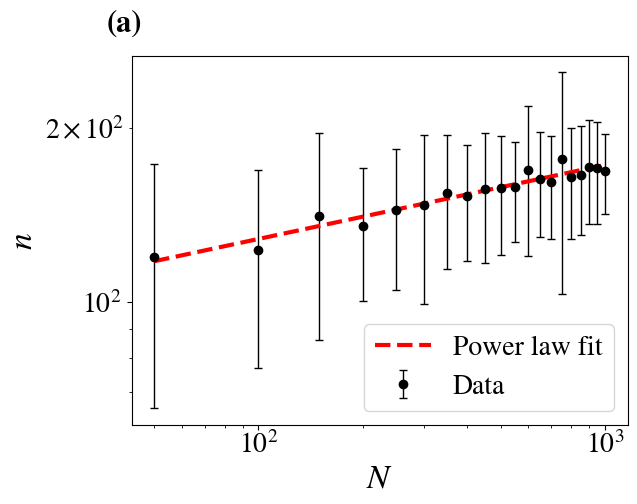}
     \end{subfigure}
     \begin{subfigure}[b]{0.24\textwidth}
         \centering
         \includegraphics[width=\textwidth]{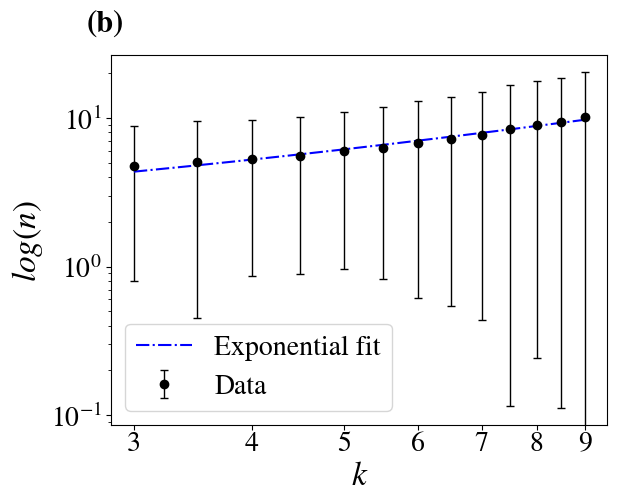}
     \end{subfigure}
     \caption{(a) \textit{$n$ vs $N$}. (b) \textit{$n$ vs $k$}. Each point is obtained from the average over 100 random graph instances with $p=0.9$. In subfigure (a), the average degree is fixed to $k=3.0$, whereas in subfigure (b) the system size is fixed to $N=50$. 
Panel (a) is shown in log–log scale, and the scaling of $n$ with $N$ is best captured by a power–law fit. In contrast, in panel (b) the best fit is exponential, which becomes a straight line when plotting $\log(n)$ on the vertical axis (see main text for fit parameters).
}
     \label{fig:scaling_classical}
\end{figure}

For the \textit{$n$–vs–$N$} scaling, the average node degree was fixed at $k = 3.0$, 
while for the \textit{$n$–vs–$k$} scaling, the number of nodes was fixed at $N = 50$ 
(see Fig.~\ref{fig:scaling_classical}). In both cases, the activation probability 
was set to $p = 0.9$.

To quantitatively characterize these trends, we fitted the data using two candidate 
functional forms: a polynomial law and an exponential law. When $k$ is fixed and $N$ 
varies (Fig.~\ref{fig:scaling_classical}a), the polynomial form $n = \gamma N^{\delta}$ 
provides the best fit, yielding the parameters $\gamma = 71.2(9)$ and $\delta = 0.12(8)$ 
with RMSE $= 3.98(8)$. We observed a sublinear scaling, indicating that the algorithm's 
convergence steps remains small as the system size grows, even up to $N = 1000$.
Conversely, when $N$ is fixed and $k$ varies (Fig.~\ref{fig:scaling_classical}b), 
the exponential model $n = e^{\Gamma 
+ \Delta k}$ captures the behavior more 
accurately. Linearizing this expression gives $\log n = \Gamma + \Delta k$, and the 
corresponding fit yields $\Gamma = 1.66(3)$ and $\Delta = 0.89(6)$, with RMSE $= 0.23(8)$. \\

These results show that the viability of the protocol strongly depends on the graph density. 
For sparse graphs (small $k$), the MIS convergence probability $P_{\mathrm{MIS}}$ remains appreciably high and can be further enhanced by tuning the activation parameter $p$, achieving a reasonable compromise between success probability and convergence time. 
As the connectivity increases, however, $P_{\mathrm{MIS}}$ decreases rapidly, reflecting the proliferation of maximal independent sets that divert probability away from the optimal ones. 
In this dense regime, maintaining a fixed target success probability would require an increasingly large number of independent runs, leading to an overall computational cost that grows exponentially with the system size.


\section{Quantum simulation}
\label{sec:quantum_simulation}

We numerically assess the performance of the quantum dissipative protocol introduced in Sec.~\ref{sec:QCA}, starting from the vacuum configuration $\ket{0}^{\otimes N}$.  
The dynamics was simulated by evolving the density operator in its vectorized Liouville-space representation, using the time-dependent variational principle (TDVP) applied to matrix product states~\cite{westhoff2025tensor} with bond dimension $\chi = 200$, as implemented in the ITensor library~\cite{itensor}. 

We first explored graphs of increasing size~$N$ and different average connectivities~$k$, focusing on the convergence behaviour of the \emph{dissipative stage} $e^{\hat{\mathcal{L}}_1 t}
|0\rangle\rangle^{\otimes N}$ described in Sec.~\ref{sec:QCA}.  
In particular, we monitored the relaxation time~$T$ required for the system to reach its steady state, that is, a statistical mixture of all maximal independent sets, as a function of both~$N$ and~$k$.  
The set of graphs considered spans system sizes in the range $N \in [10,30]$ and average degrees $k \in [1.5,5.0]$, with increments $\Delta N = 1$ and $\Delta k = 0.5$.  
For each pair $(N,k)$, the quantum dissipative evolution was integrated via TDVP until convergence within the variational manifold was achieved.  
Convergence was assessed through the overlap between consecutive states, $|\rho(t)\rangle\rangle$ and $|\rho(t+\delta t)\rangle\rangle$, obtained after each TDVP integration step of duration $\delta t=0.01$.  
The overlap was computed according to the scalar product defined in Eq.~\eqref{scalar_product}, and the simulation was terminated once the relative variation between consecutive states satisfied
\begin{equation}
1 - \frac{|\langle\!\langle \rho(t) | \rho(t+\delta t) \rangle\!\rangle|}{\langle\!\langle \rho(t) | \rho(t) \rangle\!\rangle} < 10^{-5},
\end{equation}
ensuring that the TDVP trajectory had effectively reached a stationary point of the dissipative evolution. The mean convergence time $\langle T \rangle$ and its variance were computed over 10 independent graph realizations for each $(N,k)$ pair.

\begin{figure*}[t]
    \centering
    \includegraphics[scale=0.264]{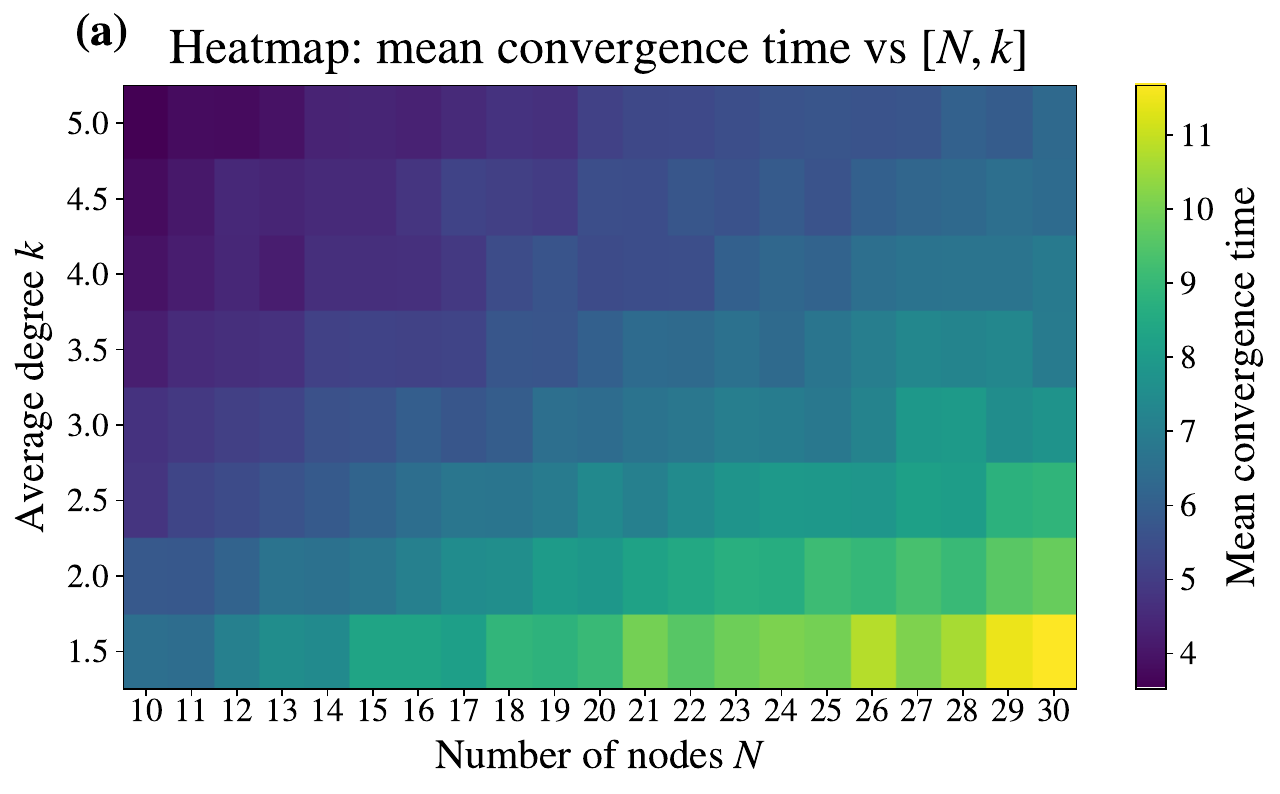}
        \includegraphics[scale=0.264]{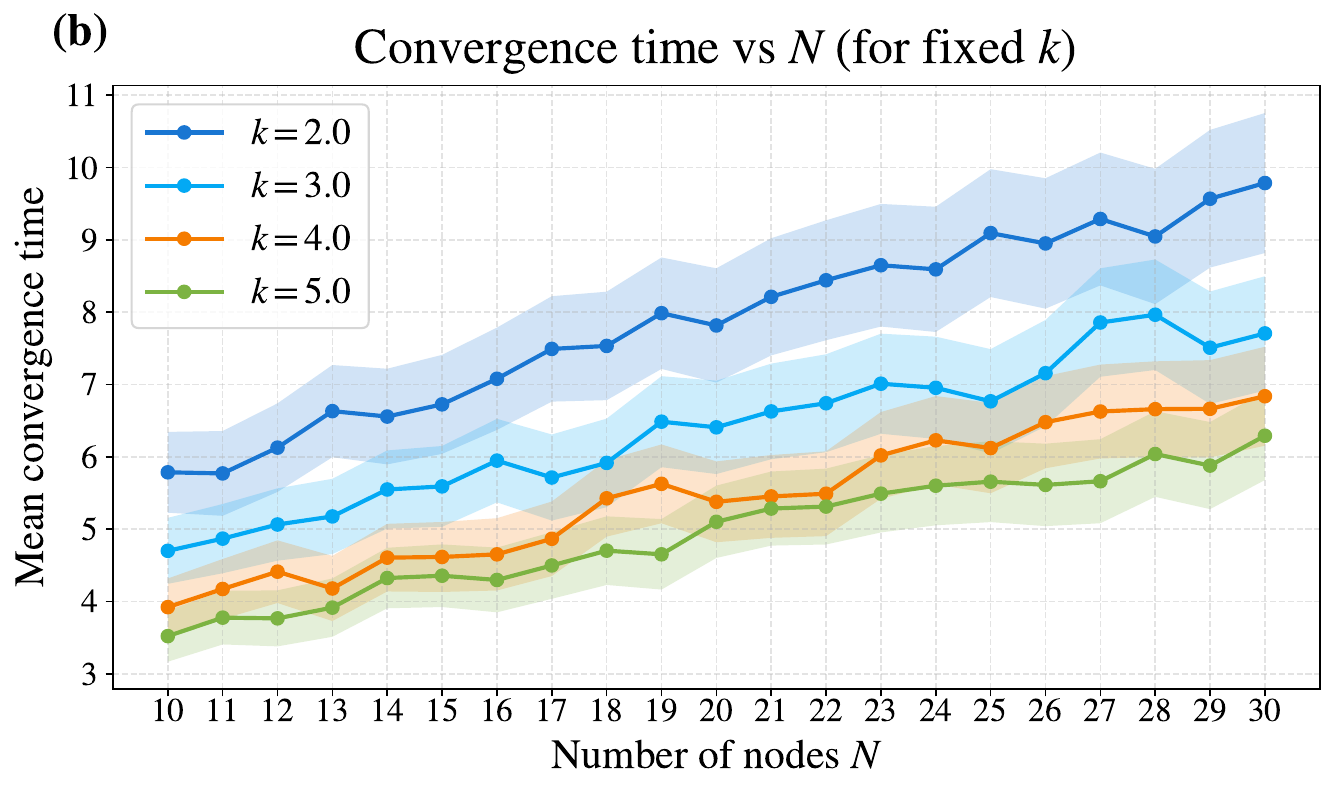}
            \includegraphics[scale=0.264]{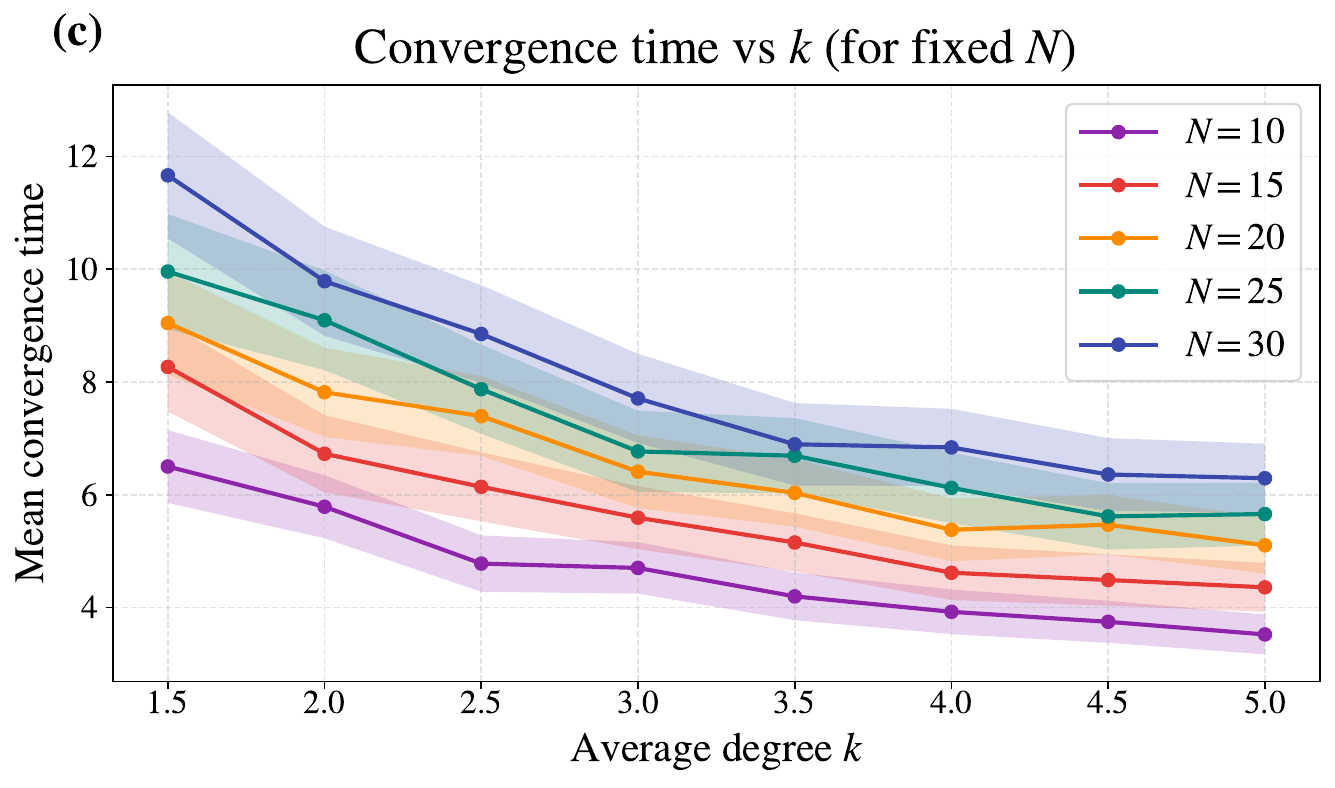}
    \caption{
    (a) Heatmap of the mean convergence time $\langle T \rangle$ for system sizes $N \in [10,30]$ and average degrees $k \in [1.5,5.0]$.
    (b) Dependence of $\langle T \rangle$ on $N$ at fixed $k$ (colored lines) with $\pm 1\sigma$ confidence bands.
    (c) Dependence of $\langle T \rangle$ on $k$ at fixed $N$.
    }
    \label{fig:quantum_triptic}
\end{figure*}

The obtained data are summarized in Fig.~\ref{fig:quantum_triptic}, which presents three complementary visualizations of the same dataset. 
Panel~(a) displays the full two-dimensional heatmap of the mean convergence time, illustrating the smooth growth of $\langle T \rangle$ with system size and its suppression at larger connectivities. 
Panels~(b) and~(c) show, respectively, the scaling of $\langle T \rangle$ as a function of $N$ at fixed $k$ and as a function of $k$ at fixed $N$, including the one-standard-deviation confidence bands obtained from the variance $\mathrm{\sigma}(T)$. 
All data follow a remarkably regular dependence on $N$ and $k$, suggesting an effective power-law scaling of the form
\begin{equation}
T_{\mathrm{fit}}(N,k) = \alpha \left( \frac{N}{k} \right)^{\beta},
\label{eq:fit_model}
\end{equation}
which captures the combined influence of the system size and graph connectivity on the convergence rate of the dissipative process.

\begin{figure}[H]
    \centering
\includegraphics[width=0.7\columnwidth]{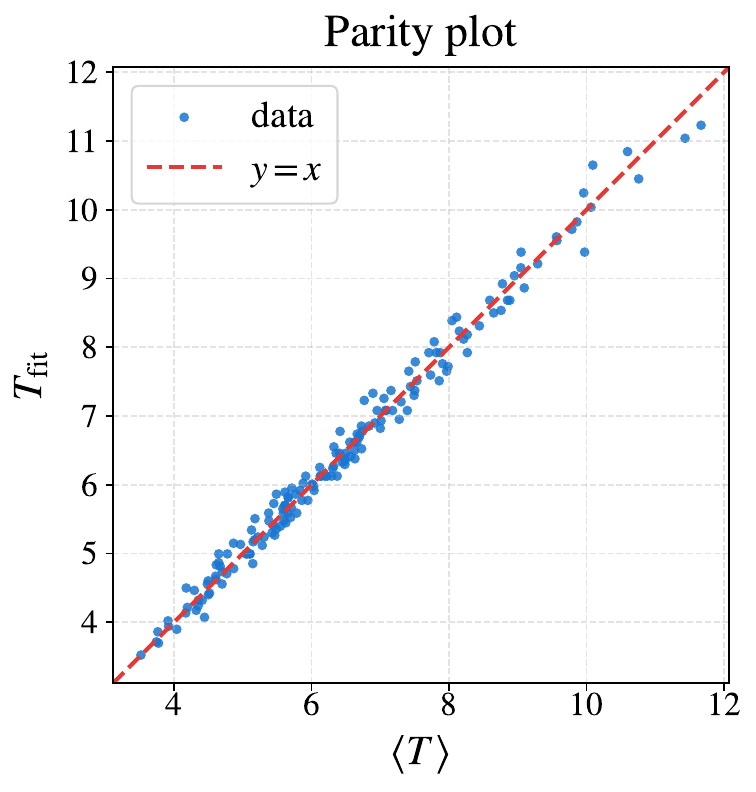}
   \caption{
Parity plot comparing the observed convergence times $\langle T \rangle$ with the fitted predictions $T_{\text{fit}} = \alpha (N/k)^{\beta}$.  
Each marker is a pair $(\langle T \rangle, T_{\text{fit}})$.  
The red dashed line indicates the ideal case $T_{\text{fit}} = \langle T \rangle$, while the strong clustering of points along the diagonal confirms the accuracy of the power-law scaling relation in Eq.~\eqref{eq:fit_model}.
}
    \label{fig:parity_plot}
    
\end{figure}

To quantitatively assess this scaling, we performed a log-linear regression of Eq.~\eqref{eq:fit_model} over the simulated dataset. 
The resulting best-fit parameters are:
$\alpha = 2.48(6)$ and $\beta = 0.50(3)$,
with a root-mean-square deviation $\mathrm{RMSE} = 0.18$. 
The strong linear correlation observed in the parity plot of Fig.~\ref{fig:parity_plot} confirms that the empirical model of Eq.~\eqref{eq:fit_model} accurately reproduces the full dataset across the explored parameter range.\\

\begin{figure}[t]
    \centering
    \includegraphics[width=0.999\columnwidth]{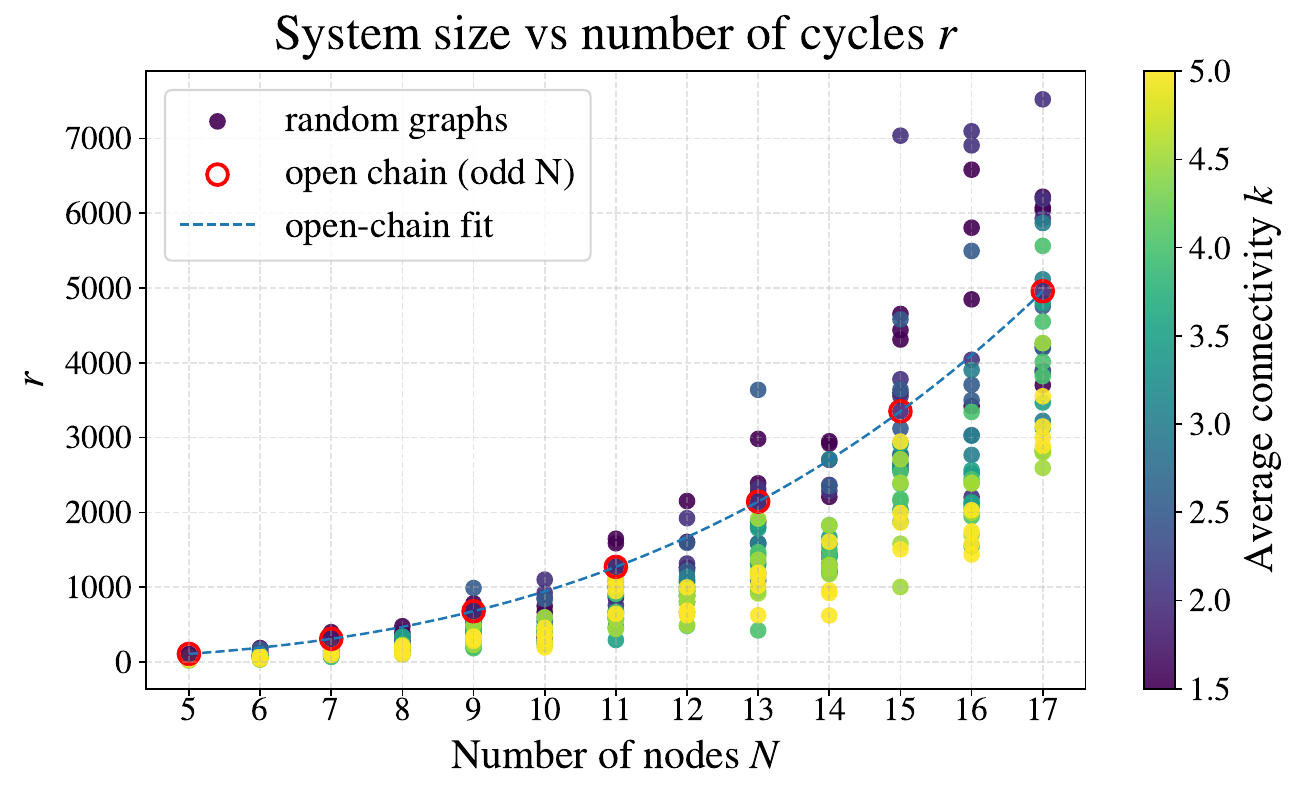}
    \caption{
    System size dependence of the number of QCA cycles required to reach a probability $P_\mathrm{MIS}>0.7$ on the optimal configuration(s).  
    Dots correspond to random unit-disk graphs colored by their average connectivity~$k$, while red circles indicate the open-chain data (odd~$N$).  
    The dashed line shows the polynomial fit $r_{\text{fit}}(N)=a N^b$ with $a=0.70(8)$ and $b=3.12(1)$.  
    The inset color bar highlights the effect of connectivity on the convergence speed.}
    \label{fig:N_vs_cycles_by_k}
\end{figure}

After establishing the scaling law of the dissipative relaxation time, we now analyze the number $r$ of cycles required to surpass a target probability on the optimal configuration (see Eq.~\eqref{eq:full_evolution}).  
Figure~\ref{fig:N_vs_cycles_by_k} summarizes the results obtained for random graphs with average connectivity $k \in [1.5,5.0]$ (step~$\Delta k = 0.5$) and system sizes $N \in [5,17]$ (step~$\Delta N = 1$).  
For $N=5$, simulations were performed only up to $k=4.0$, since higher connectivities are not allowed.
For each pair $(N,k)$, four independent graph realizations were generated, and the protocol was iterated until the probability of the maximum independent set exceeded~$0.7$.  
The parameters $t$ and $\theta$ were chosen to ensure both full relaxation of the dissipative stage and convergence of the subsequent unitary step: specifically, $t$ was taken sufficiently long to reach the steady state of $\mathcal{L}_1$—as deduced from the relaxation analysis discussed above (see Eq.~\ref{eq:fit_model})—while $\theta = 0.1$, heuristically determined, was small enough to guarantee that the target probability could be surpassed within a finite number of cycles.

To characterize the scaling behavior of the number of cycles, we did not perform an averaging over the randomly generated graph instances. Although this would be the most natural statistical procedure, the resulting data did not yield a reliable global fit nor a clear functional scaling trend. We attribute this limitation to two concomitant factors: first, the strong dependence of $r$ on the specific graph topology, which induces significant fluctuations across instances even at fixed $(N,k)$; and second, the insufficient instance sampling, due to computational restrictions, which prevented us from reaching a regime where averaging suppresses these topology-dependent deviations.
For this reason, instead of reporting averaged scalings, we opted to display the full distribution of data points in Fig.~\ref{fig:N_vs_cycles_by_k}, which provides a more faithful picture of the variability of $r$ across graph realizations. 

As a complementary and more controlled benchmark, we also analyzed a representative topology that seems to capture the global qualitative behavior: an open chain with an odd number of sites. This simple yet nontrivial structure exhibits a unique maximum independent set in the Néel-like configuration with alternating excitations and occupied boundaries, while the number of locally stable maximal configurations grows rapidly with $N$. In this setting, the QCA protocol must progressively concentrate probability weight from the manifold of maximal configurations onto the unique optimal one, making the open chain an instructive testbed for isolating the scaling properties free from instance-induced fluctuations.
In Appendix \ref{app:quantum_open_chain}, we present an analytic treatment for the smallest cases $N=3$ and $N=5$, showing explicitly how the population of the optimal state evolves during the cycles.

The open-chain data, highlighted by red circles in Fig.~\ref{fig:N_vs_cycles_by_k}, exhibit a clear polynomial scaling with $N$, well captured by the fit
\begin{equation}
r_{\text{fit}}(N) = a\,N^b, 
\qquad
a = 0.70(8),\quad b = 3.12(1),
\end{equation}
as shown by the dashed blue curve, with a root-mean-square deviation $\mathrm{RMSE} = 0.22$. 
This scaling indicates that, even though the probability flow is confined within the manifold of maximal configurations, the number of alternations between dissipative and unitary steps required to reach a fixed target probability grows polynomially with the system size.  

Importantly, the same qualitative trend is observed across random graphs: the distribution of cycle counts broadens with $N$, while the average appears to follow a similar polynomial-like dependence on $N$, with fluctuations associated with connectivity~$k$: graphs with a higher average degree tend to converge faster, as they possess fewer maximal configurations and, therefore, a reduced configuration space to explore.  
In contrast, sparse graphs (low~$k$) present a richer manifold of nearly degenerate maximal states, leading to larger variability and longer number of cycles.

\begin{figure}[h]
    \centering
    \includegraphics[width=0.99\columnwidth]{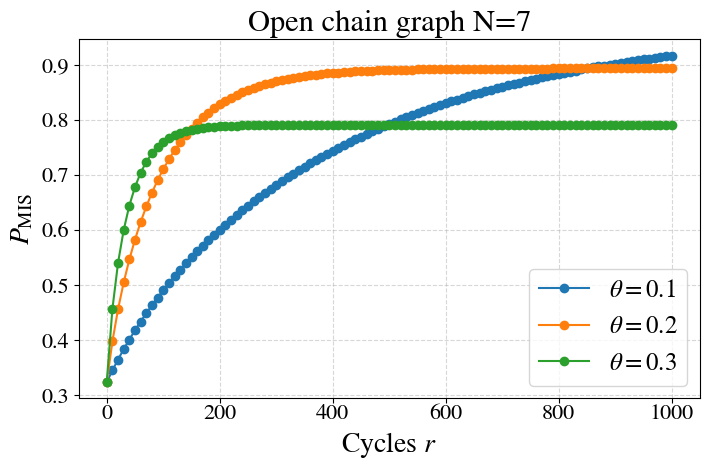}
    \caption{
    Probability of the unique MIS as a function of the number of cycles $r$ for an open chain with $N=7$ sites at different values of $\theta = [0.1,0.2,0.3]$.  
    Smaller $\theta$ lead to slower but higher asymptotic plateau, whereas larger values yield faster convergence to lower stationary values.
    }
    \label{fig:openchain_theta}
\end{figure}

Moreover, the analysis of open chains enables a clearer understanding of the role played by the parameter~$\theta$ in the protocol.
In particular, the asymptotic ($r \to \infty$) probability of the MIS configuration(s), for a fixed $\theta$, does not reach unity and follows the functional dependence
\begin{equation}
P_{\text{MIS}}(\theta) = 1 - \frac{\theta^2}{\mathcal{F}(N)\,\theta^2 + \mathcal{G}(N)},
\label{eq:open_chain_quantum}
\end{equation}
where $\mathcal{F}(N)$ and $\mathcal{G}(N)$ are $N$–dependent coefficients (see Appendix \ref{app:quantum_open_chain}). As is evident from Eq.~\eqref{eq:open_chain_quantum}, the probability approaches one only in the limit $\theta \to 0$.
This dependence on the unitary evolution parameter is explicitly illustrated in Fig.~\ref{fig:openchain_theta}, which reports the evolution of $P_{\mathrm{MIS}}$ for an open chain of $N=7$ sites and different values of $\theta$.  
After an initial rapid increase, the probability saturates to an asymptotic plateau whose value decreases monotonically with increasing~$\theta$.
Smaller values of~$\theta$ result in a slower but more pronounced convergence toward the MIS, approaching unity in the limit $\theta \to 0$, whereas larger~$\theta$ produce a faster rise followed by a lower asymptotic probability.  
Put differently, the plot highlights the intrinsic trade-off between convergence speed and asymptotic accuracy: small~$\theta$ values yield higher limiting probabilities but require a larger number of QCA cycles to reach convergence.  
From this behavior, one infers that, once a desired threshold for $P_{\mathrm{MIS}}$ is fixed, an intermediate (i.e., optimal) value of $\theta$ exists that minimizes the number of cycles needed to surpass it.

A natural question concerns how the number of cycles $r$ scales with the system size when, instead of fixing $\theta$, one chooses at each $N$ the optimal rotation angle $\theta_{\mathrm{opt}}(N)$ that minimizes the number of QCA cycles required to reach the target probability. 
The analysis presented in Fig.~\ref{fig:N_vs_cycles_by_k} is performed at a fixed value of $\theta$, which guarantees that the threshold $P_{\mathrm{MIS}}>0.7$ can be reached for all system sizes considered. 
In this regime we observe a clear polynomial growth $r\sim N^b$. However, this choice of $\theta$ would have to be decreased to preserve convergence as the system size increases, in line with the trade-off discussed above.

This fixed–$\theta$ curve should be regarded as an \emph{upper bound}, within the simulated size range, on the number of cycles required to reach the target probability. Indeed, for small systems the chosen $\theta$ is typically far from optimal, thus overestimating the minimal number of iterations that would be obtained by tuning $\theta$ at each $N$. As the system size increases, however, the fixed value of $\theta$ approaches $\theta_{\mathrm{opt}}(N)$, and the corresponding estimate becomes increasingly accurate.
If $\theta$ were optimized at each system size, the small-$N$ points would be shifted downward, whereas the large-$N$ ones would remain essentially unchanged. This adjustment would lead to a steeper effective growth of $r_{\mathrm{opt}}(N)$ over the accessible size range: the apparent scaling exponent would increase, and the resulting behaviour might no longer be well captured by the polynomial dependence observed at fixed~$\theta$.

In other words, the fixed-$\theta$ analysis correctly captures the convergence trend for moderate and large $N$, but underestimates the effective growth rate expected when using $\theta_{\mathrm{opt}}(N)$.
Our data therefore allow us to establish the polynomial scaling observed at fixed $\theta$, but do not exclude the possibility that the optimal-$\theta$ protocol exhibits a stronger asymptotic dependence on the system size.


\section{Conclusions and outlooks}
\label{sec:conclusion}

In this work we have proposed and analysed a family of classical and quantum cellular-automaton protocols for tackling the Maximum Independent Set problem on generic graphs. 
On the classical side, we introduced a synchronous probabilistic cellular automaton whose local rule enforces the independence constraint and drives the dynamics toward the set of maximal independent sets. 
Starting from the empty configuration, the PCA defines a Markov chain whose absorbing states coincide exactly with the mIS of the graph.
Numerical simulations allowed us to characterise the probability $P_{\text{MIS}}$ of converging to a maximum independent set as a function of the activation parameter $p$, the system size $N$, and the average degree $k$, as well as to estimate the scaling of the convergence time with these graph parameters.
Our construction provides, to the best of our knowledge, a first explicit use of a synchronous PCA on general graphs aimed at generating maximal independent sets, and at interpreting it within the language of the CA/PCA community.
This opens the way to systematic improvements of the update rule by experts in probabilistic cellular automata, who may optimise or generalise the dynamics while preserving its locality and parallelism.

The introduction of this classical rule also plays a conceptual role: it furnishes a natural bridge to the quantum setting.
In the second part of the work, we used the same local constraint structure to engineer a quantum cellular automaton based on the interplay between a purely dissipative Liouvillian and a constraint-preserving unitary ``PXP-like'' evolution.
The dissipative stage fills the graph with excitations wherever allowed, relaxing the system into a statistical mixture of maximal independent sets, while the coherent step redistributes probability weight within this manifold without violating the hard-core constraint.
Tensor-network simulations based on TDVP for vectorised density matrices enabled us to quantify the scaling of the dissipative convergence time with $N$ and $k$, and to extract an empirical power-law dependence $T_{\mathrm{fit}}(N,k) = \alpha (N/k)^{\beta}$ over the range of accessible sizes.
For the full dissipative–unitary protocol, we analysed how many QCA cycles are required to boost the probability of the optimal configuration(s) above a fixed threshold, both for random graphs and for an open chain, finding an effective polynomial scaling with $N$ at fixed driving angle $\theta$.

These results suggest that local, translationally invariant QCA dynamics can provide a promising physics-inspired heuristic for MIS, complementary to Hamiltonian-based approaches.
At the same time, our study also highlights several limitations and directions for improvement.
First, the analysis of the open chain was performed at fixed $\theta$, which provides a controlled but not yet optimal estimate of the convergence time: a more accurate scaling would require, for each system size $N$, a systematic optimisation of the rotation angle $\theta_{\mathrm{opt}}(N)$, so as to minimise the number of cycles needed to reach a target $P_{\text{MIS}}$.
This optimisation is computationally demanding and was beyond the scope of the present work, but it will be crucial to establish whether the apparent power-law behaviour observed at fixed $\theta$ persists when using $\theta_{\mathrm{opt}}(N)$ or instead crosses over to a different scaling.
Second, the TDVP simulations were restricted to moderate graph sizes; extending the analysis to larger $N$ will be essential to confirm that the empirical law is not merely a small-size effect. 

A further observation concerns the robustness and universality of the quantum routine.
For the ensemble of graph instances considered here, we find that in a small fraction of cases (about $\sim 7\%$ of all instances) the protocol fails to increase $P_{\text{MIS}}$ beyond a given threshold, irrespective of the value of $\theta$ explored within our parameter range.
This behaviour indicates that, for certain graphs, the chosen PXP-type driver is not sufficiently effective at funnelling probability toward the optimal configuration(s), even when assisted by the dissipative reset.
From an algorithmic point of view, this motivates the design and analysis of alternative constraint-preserving Hamiltonians or more structured unitary layers, potentially tailored to the graph topology, which could achieve a faster and more reliable concentration of probability on MIS while requiring an equal or smaller number of dissipative–unitary cycles.

Finally, the local and translationally invariant nature of our constructions makes them naturally compatible with neutral-atom platforms and other programmable quantum simulators, offering a concrete pathway to hardware implementations of CA-inspired optimisation protocols and to systematic experimental studies of their performance on realistic noisy devices.

\section{Acknowledgments}
We thank Elisa Ercolessi for valuable guidance and Riccardo Cioli for helpful feedback and discussions.
The research of F.D. was supported by the 2022-PRIN Project “Hybrid algorithms for quantum simulators”, that of V.G. by the Project “ACAT” 
from the National Centre for HPC, Big Data, and Quantum Computing (Spoke 10, CN00000013). 
\clearpage
\clearpage
\twocolumn[
    \centering
    \section*{References}
    \vspace{1em} 
]
\balance
\bibliography{ref}
\bibliographystyle{ieeetr}
\clearpage
\twocolumn[
    \centering
    \section*{Appendix}
    \vspace{1em} 
]
\appendix
\section{$\mathbf{P}_{\text{MIS}}$ computation}
\label{app:calcolo_pmis}
\label{app:triangolopene}

\begin{figure}[h]
    \centering
\includegraphics[width=0.7\linewidth]{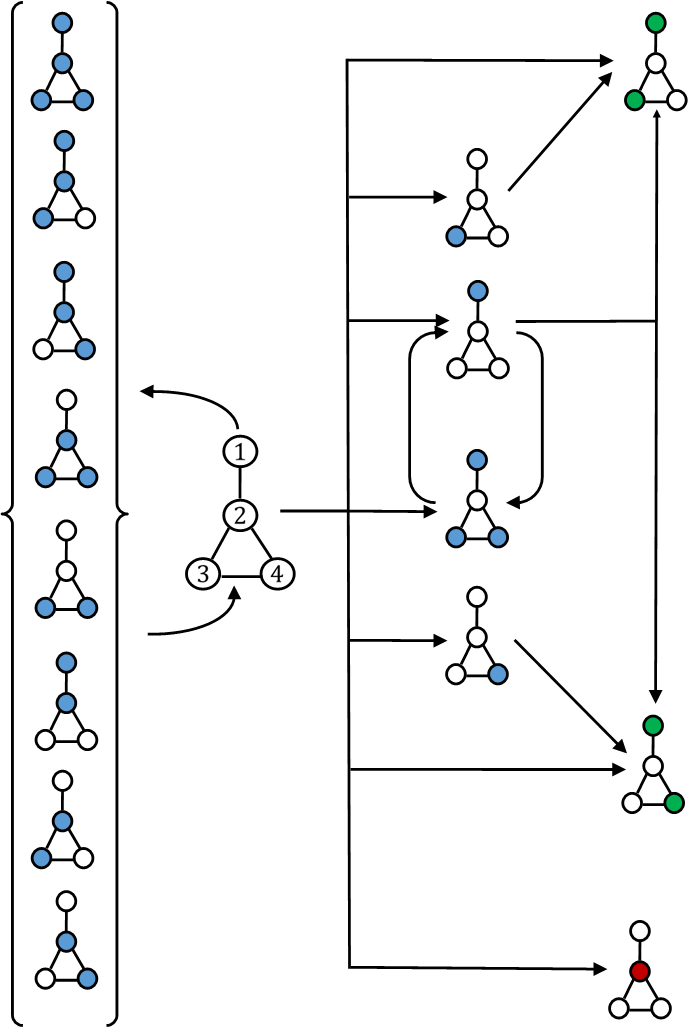}
    \caption{Transition scheme of the 4-node graph under consideration}
    \label{fig:triangolopene}
\end{figure}

In this section, we present the full derivation of the probability that the algorithm converges to a maximal independent set (MIS) for the four-node instance shown in Fig.~\ref{fig:triangolopene}, where the two MIS and the mIS are the green and red configurations, respectively. 

The graph admits 16 possible configurations, represented by the bitstrings $[0000, 0001, \ldots, 1110, 1111]$. Recalling the update rules of the algorithm and defining $q := 1 - p$, the corresponding transition scheme is reported in Table~\ref{tab:triangolopene_trans}.

Starting from the initial configuration $0000$, the PCA can evolve into any possible configuration of the graph (including the initial one itself), with transition probabilities determined by the number of nodes that become active in the final configuration. Since all nodes are initially inactive ($0$), each node can be switched on with probability $p$ or remain inactive with probability $q$. Therefore, in the first step, for any possible configuration of the graph, the transition probability is given by $p$ raised to the power of the number of active nodes (denoted $\#1$) multiplied by $q$ raised to the power of the number of inactive nodes (denoted $\#0$).

It is worth noting that most configurations transition to $0000$ with unit probability, due to the update rule enforcing the non-adjacency condition by setting to $0$ any pair of adjacent nodes that are simultaneously active (see Fig.~\ref{fig:triangolopene}).

Consider now the configurations $0010$ and $0001$. In these cases, node $3$ or node $4$, respectively, is active, and both are connected to node $2$. The triangle formed by nodes $2$–$3$–$4$ is thus stable and remains unchanged. The only node that can still change its state is node $1$, which can be activated with probability $p$ or remain inactive with probability $q$ (see Table~\ref{tab:triangolopene_trans}). After a sufficient number of iterations, node $1$ will be activated, leading the system to a stationary MIS configuration and thereby terminating the algorithm. By the same reasoning, one can understand the transition probabilities from configurations $1000$ and $1011$.

To compute the overall probability of convergence to a MIS, it is necessary to account for all possible transition paths leading to a MIS, including infinitely many paths in which the PCA revisits configuration $0000$ an arbitrary number of times. For instance, the MIS $1010$ can be reached by activating nodes $1$ and $3$ after the algorithm has “tested” several configurations that reset the system to $0000$ (hereafter referred to simply as configuration $0$). The probability of returning to one of these configurations is given by

\begin{equation}
G_0 = p^4 + 3p^3q + 4p^2q^2 + q^4.
\end{equation}

Consequently, the probability of reaching $1010$ by passing through configuration $0$ an arbitrary number of times is

\begin{equation}
P_{1010}^{partial} = \sum_{i=0}^{\infty} p^2 q^2 (G_0)^i = \frac{p^2 q^2}{1 - G_0}.
\end{equation}

This same reasoning applies to the MIS $1001$.

The key insight lies in the geometric series that arises from paths in which certain configurations are revisited an arbitrary number of times. As another example, we can reach the MIS configurations $1001$ and $1010$ by first reaching configuration $1000$ (possibly passing through $0$ several times), and then reaching one of the two MIS configurations, passing through $1011$ with probability $p^2$ or remaining in $1000$ with probability $q^2$, any number of times:

\begin{equation}
P_{1001,1010}^{via\;1000} = \frac{p^3 q + p q^3}{1 - G_0} \cdot \frac{2pq}{1 - (p^2 + q^2)}.
\end{equation}

Using the identity $1 = (p + q)^2$, we find $1 - (p^2 + q^2) = 2pq$, which cancels with the numerator. This simplification reflects the fact that, once the algorithm reaches either configuration $1000$ (with probability $p q^3$) or $1011$ (with probability $p^3 q$), it enters a loop that inevitably leads to a MIS.

By combining all possible paths to a MIS and exploiting the same simplification, $1 = (p + q)^4$, we obtain the following expression for $P_{\text{MIS}}$:

\begin{equation}
P_{\text{MIS}} = \frac{p^3 q + 2p^2 q^2 + 3p q^3}{1 - G_0} = \frac{p^3 q + 2p^2 q^2 + 3p q^3}{p^3 q + 2p^2 q^2 + 4p q^3}.
\end{equation}

In contrast, the probability of converging to a maximal independent set (mIS) is given by
\begin{equation}
P_{\text{mIS}} = \frac{p q^3}{1 - G_0}.
\end{equation}

It is straightforward to verify that $P_{\text{MIS}} + P_{\text{mIS}} = 1$, and that
\begin{equation}
\lim_{q \rightarrow 0} P_{\text{MIS}} = 1.
\end{equation}

\begin{figure}[h]
    \centering
    \centering
    \includegraphics[width=0.4\linewidth]{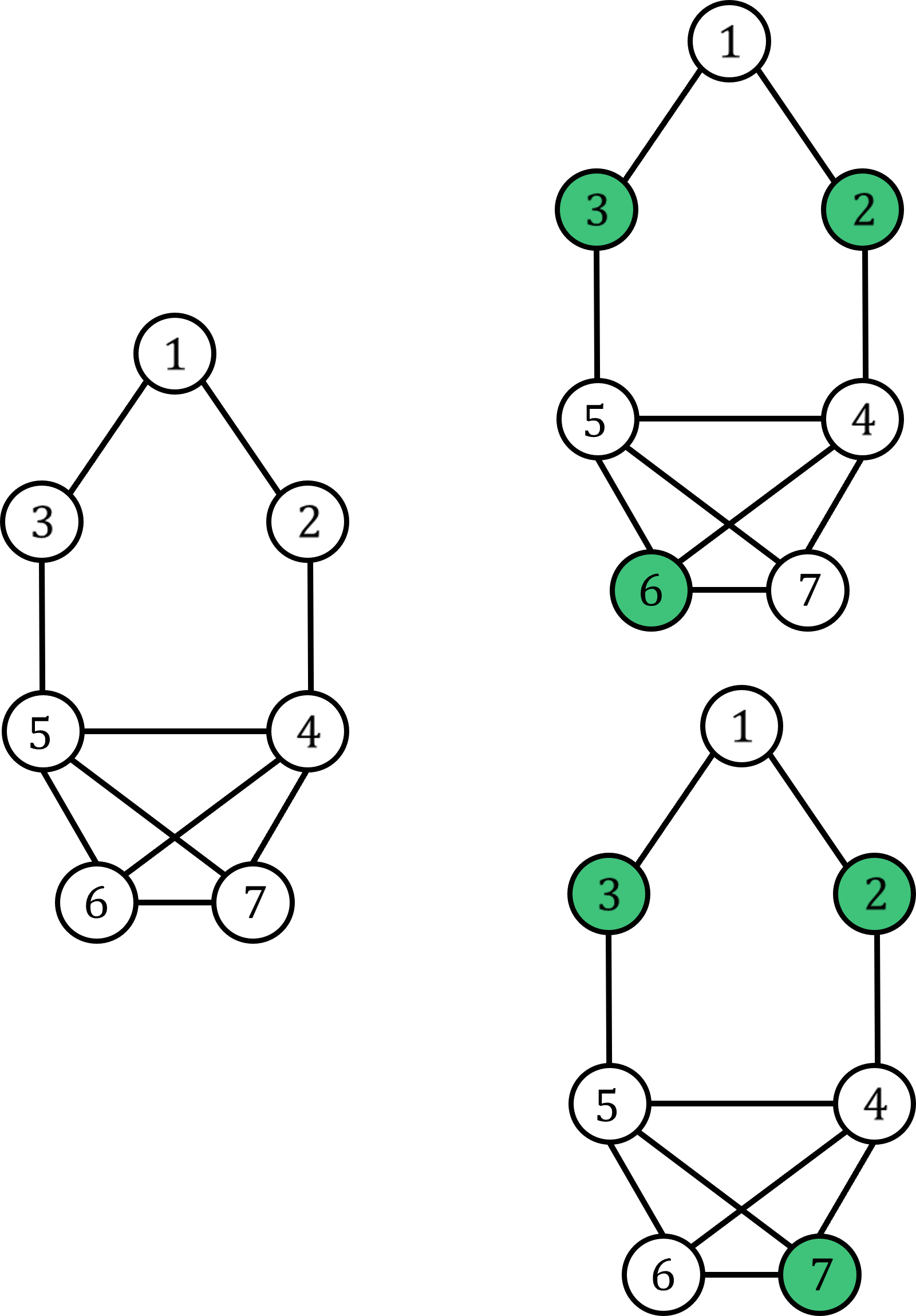}
    \caption{Pathologic case: the house graph and its Maximum Independent Sets}
    \label{fig:casetta}
\end{figure}

In the limit of large $p$, the probability of reaching a MIS is dominated by the term with the highest power of $p$ and the lowest power of $q$, which in this case is $p^3 q$. The order of the dominant term in $p$ and $q$ depends on the structure of the underlying graph (the term $p^{|G|}$, corresponding to $p^4$ in this instance, does not contribute). Importantly, the limit above depends on the coefficients of the dominant terms in both the numerator and the denominator. These coefficients are not always equal, as in certain graph instances the most probable nontrivial transition paths, those associated with configurations containing the largest number of active nodes (or, equivalently, the highest powers of $p$), do not necessarily lead to a MIS. In such cases, $P_{\text{MIS}}$ exhibits a limiting value strictly smaller than 1. An example of this behavior is provided in Appendix \ref{app:casetta}.
\begin{table}[h]
    \centering
    \begin{tabularx}{\columnwidth}{|X|X|X|}
        \hline
        \textbf{Configuration at step $i$} &
        \textbf{Possible configurations at step $i+1$} &
        \textbf{Transition probability} \\ \hline
        0000 & all & $p(\#1)\cdot q(\#0)$ (see main text) \\ \hline
        0001 & 1001 & $p$ \\
        & 0001 & $q$ \\ \hline
        0010 & 1010 & $p$ \\
        & 0010 & $q$ \\ \hline
        0011 & 0000 & 1 \\ \hline
        0100 (mIS) & End & \\ \hline
        0101 & 0000 & 1 \\ \hline
        0110 & 0000 & 1 \\ \hline
        0111 & 0000 & 1 \\ \hline
        1000 & 1011 & $p^2$ \\ 
        & 1010 & $pq$ \\
        & 1001 & $pq$ \\
        & 1000 & $q^2$ \\ \hline
        1001 (MIS) & End & \\ \hline
        1010 (MIS) & End & \\ \hline
        1011 & 1000 & 1 \\ \hline
        1100 & 0000 & 1 \\ \hline
        1101 & 0000 & 1 \\ \hline
        1110 & 0000 & 1 \\ \hline
        1111 & 0000 & 1 \\ \hline
    \end{tabularx}
    \caption{Full transition scheme for the 4-node graph under consideration}
    \label{tab:triangolopene_trans}
\end{table}
\label{app:casetta}
A rather simple pathologic case in which the convergence probability on a MIS in the limit $p\rightarrow 1$ is non-unitary is what we called the "house graph" (fig. \ref{fig:casetta}). \newline Out of 128 possible configurations the graph has 72 zero-absorbing configurations, 8 stationary states and 48 non-absorbing non-stationary configurations. By repeating the same calculations shown for the simple 4-node graph, one can find that the probability that the algorithm converges into a MIS is given by:
\begin{align}
    &P_{\text{MIS}}=\frac{2p^3q^4}{1-G_1}+2\frac{p^5q^2+2p^4q^3+2p^3q^4+pq^6}{1-G_1}\frac{2pq^3}{1-G_2}+\\ \nonumber
    &+2\frac{p^5q^2+2p^4q^3+2p^3q^4+pq^6}{1-G_1}\frac{p^3q+pq^3}{1-G_2}\frac{2pq}{1-p^2-q^2}+\\ \nonumber 
    &+2\frac{p^4q^3+2p^3q^4+pq^6}{1-G_1}\frac{p^2q}{1-G_3}+ \\ \nonumber
    &+\frac{p^4q^3+p^2q^5}{1-G_1}\frac{2pq}{1-p^2-q^2}+ \\ \nonumber
    &+2\frac{p^4q^3+2p^3q^4+pq^6}{1-G_1}\frac{2pq^2}{1-G_3}+\frac{4p^2q^5}{1-G_1} \nonumber
\end{align}
where the terms $G_1, G_2, G_3$ are the following:
\begin{align}
    G_1=&p^7+7p^6q+18p^5q^2+22p^4q^3+13p^3q^4+10p^2q^5+q^7 \\
    G_2=&p^4+3p^3q+4p^2q^2+q^4 \nonumber \\
    G_3=&p^3+2p^2q+q^3. \nonumber
\end{align}
Taking the limit $\lim_{q\rightarrow 0} P_{\text{MIS}}$ is quite simple since, in each fraction, the dominant term is the one with the lowest power of $q$. To find the correct leading order in the denominators, we rewrite $1=(p+q)^k$ where $k$ is the highest power in $G_i$ and explicitly list all the terms of the binomial power to cancel out the highest and lowest power of $q$ in the $G_i$ terms. After simplifications, one gets:
\begin{equation}
    \lim_{q\rightarrow 0}P_{\text{MIS}}=\lim_{q\rightarrow 0}\frac{4q^3}{3p^3}+\frac{10q^2}{3p^2}+\frac{q}{p}+\frac{2}{3}=\frac{2}{3}.
\end{equation}
Summarizing, although $P_{\text{MIS}}$ does not approach $1$ like in the previous case, our conjecture is still verified since each MIS has, for symmetry reasons, a probability of $1/3$ to be reached, while each mIS has a probability upper-bounded by $1/3$ (a rough estimate is given by $(1-P_{\text{MIS}})/(\#\text{mIS})=1/18$ each mIS).
\clearpage
\section{Quantum open chain}
\label{app:quantum_open_chain}
We consider an open chain composed of 3 nodes and aim to compute the full expression of the MIS convergence probability, denoted as $P_{\text{MIS}}$. In this specific case, we label it $P_{101}$, since the bitstring $101$ represents the Maximum Independent Set of the 3-node chain. The system evolves under repeated applications of the $\hat{\mathcal{L}}$–$H$ dynamics, where the Hamiltonian evolution driven by $H$ is expanded in powers of the angle $\theta$ up to fourth order ($\theta^4$). Our goals are twofold: (i) to derive an analytical expression for the critical angle $\theta_T$ such that the asymptotic value of $P_{101}$ reaches a chosen threshold $T$, and (ii) to determine the number of evolution steps required for $P_{101}$ to reach the same threshold $T$ for a fixed value of $\theta$.

The first step is to apply the vectorized Lindbladian operator onto the density matrix of the initial state:
\begin{equation}
\label{eq:lindbladian_on_000}
e^{\hat{\mathcal{L}}t} |000\rangle\rangle = P_{101}^0 |101\rangle\rangle + P_{010}^0 |010\rangle\rangle
\end{equation}
where we have introduced the shortcut
\begin{align}
|101\rangle\rangle := |101\rangle\langle 101| \\
|010\rangle\rangle := |010\rangle\langle 010|\nonumber \\
P_{101}^0 = \frac{2}{3}, \quad P_{010}^0 = \frac{1}{3}\nonumber
\end{align}
The Hamiltonian is given by:
\begin{equation}
H = \sum_i^3 P_{i-1} X_i P_{i+1}
\end{equation}
and expanding its exponential in $\theta$ results in
\begin{equation}
 U(\theta)=e^{-i\theta H} \simeq 1 - i\theta H - \frac{\theta^2 H^2}{2} + i\frac{\theta^3 H^3}{3!} + \frac{\theta^4 H^4}{4!} + \mathcal{O}(\theta^5).
\end{equation}
The action of the unitary evolution on the state $|101\rangle$ is the following:
\begin{align}
&U(\theta) |101\rangle = |101\rangle - i\theta \biggl[|100\rangle + |001\rangle\biggl] - \\
&-\frac{\theta^2}{2}\biggl[2|000\rangle + 2|101\rangle\biggl] + i\frac{\theta^3}{3!}\biggl[4|100\rangle + 4|001\rangle + 2|010\rangle\biggl] +\nonumber \\
&+\frac{\theta^4}{4!}\biggl[10|000\rangle + 8|101\rangle\biggl],\nonumber
\end{align}
resulting in the following density matrix:
\begin{align}
&U(\theta) |101\rangle\langle 101| U^\dagger(\theta) = \biggl(1 - \theta^2 + \frac{\theta^4}{3}\biggl)^2 |101\rangle \langle 101| + \\
&+\biggl(\theta^2 - \frac{4\theta^4}{3}\biggl)\biggl[|100\rangle\langle 100| + |001\rangle\langle 001|\biggl] + \theta^4|000\rangle\langle 000|.\nonumber
\end{align}
On the other hand, the action of the Hamiltonian on the state $|010\rangle$, corresponding to the mIS of the graph is:
\begin{align}
&U(\theta) |010\rangle = |010\rangle - i\theta |000\rangle - \frac{\theta^2}{2}\biggl[|100\rangle + |010\rangle + |001\rangle\biggl] + \\ 
&+ i\frac{\theta^3}{3!}\biggl[3|000\rangle + 2|101\rangle\biggl] + \frac{\theta^4}{4!}\biggl[5\biggl(|100\rangle + |001\rangle\biggl)+ 3|010\rangle\biggl],\nonumber
\end{align}
and the density matrix comes out to be:
\begin{align}
&U(\theta) |010\rangle\langle 010| U^\dagger(\theta) = \biggl(1 - \frac{\theta^2}{2} + \frac{\theta^4}{8}\biggl)^2 |010\rangle \langle 010| + \\
&+\frac{\theta^4}{4}\biggl[|100\rangle\langle 100| + |001\rangle\langle 001|\biggl] + (\theta^2-\theta^4)|000\rangle\langle 000|.\nonumber
\end{align}

By defining:
\begin{equation}
[\hat{\mathcal{L}}]:=e^{\hat{\mathcal{L}}t} \biggl[U(\theta)\biggl(e^{\hat{\mathcal{L}}t} |000\rangle\rangle\biggl)U^{\dagger}(\theta)\biggl]
\end{equation}
indicating the operator performing a full step starting from the initial state, we can compute the action of the full operator onto the two fundamental states:
\begin{equation}
[\hat{\mathcal{L}}] |101\rangle\rangle = \biggl(1 - \frac{1}{3}\theta^4\biggl) |101\rangle\rangle + \frac{\theta^4}{3} |010\rangle\rangle;
\end{equation}
\begin{equation}
[\hat{\mathcal{L}}]|010\rangle\rangle = \biggl(1 - \frac{2}{3}\theta^2 + \frac{1}{6}\theta^4\biggl) |010\rangle\rangle + \left(\frac{2}{3}\theta^2 - \frac{1}{6}\theta^4\right) |101\rangle\rangle.
\end{equation}
This can be recovered by recalling the action of the Lindbladian onto the non-mIS states. In particular, the states $|001\rangle\rangle,|100\rangle\rangle$ will be dissipated to the MIS $|101\rangle\rangle$ while the null state $|000\rangle\rangle$, as stated in eq. \ref{eq:lindbladian_on_000}, is splitted into both $|101\rangle\rangle$ and $|010\rangle\rangle$ with components $2/3$ and $1/3$ respectively.
The state of the system at $r=1$, where $r$ indicates the number of steps is then:
\begin{align}
&|\psi\rangle\rangle := P_{101}^1 |101\rangle\rangle + P_{010}^1 |010\rangle\rangle
\end{align}
and the probability to measure the MIS, i.e. the component of $|101\rangle\rangle$ is:
\begin{equation}
    P^1_{101}=P^0_{101}\biggl(1-\frac{1}{3}\theta^4\biggl)+P^0_{010}\biggl(\frac{2}{3}\theta^2-\frac{1}{6}\theta^4\biggl),
\end{equation}
which can be rewritten as:
\begin{equation}
P_{101}^1 = P_{101}^0 \biggl(1 - \frac{2}{3}\theta^2 + \frac{1}{6}\theta^4\biggl) + \frac{2}{3}\theta^2 - \frac{1}{6}\theta^4.
\end{equation}
Let's define $K$ and $Z$ as:
\begin{equation}
K = 1 - \frac{2}{3}\theta^2 - \frac{1}{6}\theta^4, \qquad
Z = \frac{2}{3}\theta^2 - \frac{1}{6}\theta^4,
\end{equation}
then
\begin{equation}
P_{101}^1 = P_{101}^0 K + Z.
\end{equation}
After $r$ evolution steps, the MIS probability $P_{101}^r$ comes out to be:
\begin{align}
&P_{101}^r = P_{101}^0 K^r + Z\sum_{i=0}^{r-1} K^i = \\
&=P_{101}^0 K^r + Z \frac{1 - K^r}{1 - K}\nonumber
\end{align}
Increasing the number of steps, the MIS probability converges to a value which can be computed as:
\begin{align}
&P_{101}^{\infty}:=\lim_{r \to \infty} P_{101}^r = \frac{Z}{1 - K} = \\
&=1-\frac{\theta^2}{\frac{3}{4}\theta^2+1}.\nonumber
\end{align}
Now, let's set a threshold $T$ such that $P_{101}^{\infty}>T$, the condition on the angle $\theta$ to satisfy this constraint is found as:
\begin{align}
1-\frac{\theta_T^2}{\frac{3}{4}\theta_T^4 + 1} > T
\Rightarrow \theta_T < \sqrt{\frac{4(1-T)}{4 + 3T}}
\end{align}
with the weaker condition $T > -\frac{1}{3}$ for the existence of $\theta_T$ and a stronger condition $T > \frac{3}{7} $ in order $\theta_T<1$, validating the series expansion.

Now, suppose we choose a threshold $T$ such that $P_{101}^r>T$ and an angle $\theta$ smaller than $\theta_T$, the number of steps required to achieve such condition is computed as:
\begin{align}
&P_{101}^0 K^r + Z\frac{1 - K}{1 - K^r} > T \Rightarrow \\
&\Rightarrow r > \log_K \left( \frac{T - P^{\infty}_{101}}{P_{101}^0 - P^{\infty}_{101}} \right).\nonumber
\end{align}
Experimentally, we have checked that the angle $\theta$ to achieve the above condition with the smallest number of steps is just slightly smaller than the critical angle $\theta_T$. By choosing
\begin{equation}
\theta = \sqrt{ \frac{4(1-T)}{1+3T} } - \delta, \quad \delta \ll \theta,
\end{equation}
and defining
\begin{equation}
M := \sqrt{\frac{4(1-T)}{1 + 3T}},
\end{equation}
we can find
\begin{equation}
r > \log_K \left( \frac{T + \frac{(M^2 - 2\delta M)}{\frac{3}{4}(M^2 - 2\delta M) + 1} + 1}{P^0_{101}+\frac{(M^2 - 2\delta M)}{\frac{3}{4}(M^2 - 2\delta M) + 1}-1} \right)
\end{equation}
where we disregarded terms $\mathcal{O}(\delta^2)$. The logarithm on the right-hand side has the smallest possible value.

\paragraph{The 5-node chain}
For the sake of completeness, and to justify the ansatz in eq. \ref{eq:open_chain_quantum}, we report that, for the 5-node chain, the value $P^{\infty}_{10101}$, where $10101$ corresponds to the 5-node MIS, is:
\begin{equation}
    P^{\infty}_{10101}=1-\frac{\theta^2}{\frac{5}{16}+\frac{147}{160}\theta^2}.
\end{equation}
\section{QCA on Rydberg atoms}
\label{app:rydberg}
Quantum Cellular Automata can be implemented with arrays of Rydberg atoms by exploiting native Van Der Waals interactions, scaling as a low power of the distance, namely $V\propto1/R^6$, and the possibility of adding non-trivial dissipative interaction terms. Furthermore, the three-level dynamics of the system can be mapped into a two-level one by adding an extra dissipative term in the effective time-independent two-level Hamiltonian including the decay rate $1/\Gamma$ from the intermediate state to the ground. This transformation is methodically shown in \cite{Unitary_Wintermantel} for 1D systems. In this paper we promise to extend the derivation of the two-level dynamics for 2D systems with the aim of expanding the application of QCA on Rydberg atoms to more complex graph problems.

Rydberg atoms can be modeled as made out of three main electronic levels, namely the ground state, which is $|g\rangle=|5S_{1/2}\rangle$ for Rubidium, one of the most common atomic species employed in neutral atom platforms, an intermediate state $|e\rangle=|6P_{3/2}\rangle$ and an highly excited Rydberg state $|r\rangle=|nS_{1/2}\rangle$ \cite{Unitary_Wintermantel}. Let's call $V$ the coupling between nearest neighbors (for simplicity we restrict ourselves to the case where the distance between NN is the same for each atom couple). Two different multifrequency fields couple the transitions $|g\rangle\rightarrow|r\rangle$ and $|e\rangle\rightarrow|r\rangle$, respectively. Let's label the two fields $\theta_j^k$ and $\phi_j^k$ acting on site $j$ with laser detuning $kV$ with respect to the resonant frequency, where $k\in[0,K-1]$. In 1D, the Hamiltonian of the system is 
 \begin{equation}
    H=\sum_{j,k}\Bigl(\frac{\theta_j^k}{2}e^{ikVt}\sigma^{gr}_j+\frac{\phi_j^k}{2}e^{ikVt}\sigma^{er}_j+h.c.\Bigl)+V\sigma^{rr}_j\sigma^{rr}_{j+1}
\end{equation}
where $\sigma^{ab}=|a\rangle\langle b|$.

Following the scheme proposed in \cite{Unitary_Wintermantel} where the time-dependent Hamiltonian is rewritten in the rotating frame, we move to the interaction picture by performing the unitary transformation $U=\text{exp}(-\frac{iVt}{2}\sum_p\sum_{j_p}V\sigma^{rr}_p\sigma^{rr}_j)$ where we label with $j_p$ the set of nearest neighbors of site $p$. The factor $1/2$ in front of the sum avoids double counting. The Pauli operator $\sigma^{\alpha r}_p$ therefore becomes $\sigma^{\alpha r}_p=\sum_{j_p}\Bigl(P^0_j+P^1_je^{-iVt}\Bigl)$ with $P^0_j=1-\sigma^{rr}_j$ and $P^1_j=\sigma^{rr}_j$. \newline The time-independent Hamiltonian can be rewritten as:
\begin{align}
    & H'=\frac{1}{2}\sum_p\sum_{\alpha=0}^1\prod_{j_p}P^{\alpha_j}_j\Bigl(\theta^k_p\sigma^{gr}_p+\phi^k_p\sigma^{er}_p+h.c.\Bigl) \\
    & k=\sum_{j_p}\alpha_j
\end{align}
The sum over $\alpha$ and the superscript $\alpha_j$ mean that the type of projector is different for each nearest neighbor $j\in\{j_p\}$ and we sum over all the possible combinations of projector products. The total number $K$ of frequencies addressing the same transition must therefore be $K\geq \max_p(||\{j_p\}||)$, where $||\{j_p\}||$ indicates the cardinality of the ensemble of nearest neighbors. \newline The spontaneous decay from the intermediate state is included in the jump operators $L_p=\sqrt{\Gamma}\sigma^{ge}_p$. To eliminate the manifold of the intermediate states, we move to the effective operator formalism. It can be proven that the effective Hamiltonian and the effective jump operators are:
\begin{align}
    & H_{eff}=\frac{1}{2}\sum_p\sum_{\alpha=0}^1\prod_{j_p}P^{\alpha_j}_j\Bigl(\theta^k_p\sigma^{gr}_p+h.c.\Bigl) \\
    & L^{eff}_p=\frac{i}{\sqrt{\Gamma}}\sum_{\alpha=0}^1\prod_{j_p}P^{\alpha_j}_j\phi^k_p\sigma^{gr}_p.
\end{align}
The dynamics of the system can be recovered by solving the master equation
\begin{align}
    &\partial_t\rho=\mathcal{L}(\rho)=-i[H_{eff},\rho]+ \\
    &+\sum_pL^{eff}_p\rho (L^{eff}_p)^{\dagger}-\frac{1}{2}\{L^{eff}_p)^{\dagger}L^{eff}_p,\rho\}. \nonumber
\end{align}
This whole framework is valid under the assumption \\ $V\gg\Gamma>\theta^k_p,\phi^k_p$.



\end{document}